\documentclass[conference]{IEEEtran}
\IEEEoverridecommandlockouts
\usepackage{cite}
\usepackage{amsmath,amssymb,amsfonts}
\usepackage{algorithm}
\usepackage{algorithmic}
\usepackage{graphicx}
\usepackage{textcomp}
\usepackage{xcolor}
\usepackage{subcaption}
\usepackage{multirow}

\def\BibTeX{{\rm B\kern-.05em{\sc i\kern-.025em b}\kern-.08em
    T\kern-.1667em\lower.7ex\hbox{E}\kern-.125emX}}
 
\providecommand{\parab}[1]{\noindent\textbf{#1}}

\begin{document}

\title{Joint Optimization of Signal Design and Resource Allocation in Wireless D2D Edge Computing
}

\author{\IEEEauthorblockN{Junghoon Kim\IEEEauthorrefmark{1}, Taejoon Kim\IEEEauthorrefmark{2}, Morteza Hashemi\IEEEauthorrefmark{2}, Christopher G. Brinton\IEEEauthorrefmark{1}, and David J. Love\IEEEauthorrefmark{1} }
\IEEEauthorblockA{\IEEEauthorrefmark{1}Electrical and Computer Engineering, Purdue University, IN, USA
\\
\IEEEauthorrefmark{2}Electrical Engineering and Computer Science, University of Kansas, KS, USA \\
Email: \IEEEauthorrefmark{1}{\{kim3220, cgb, djlove\}}@purdue.edu,
\IEEEauthorrefmark{2}{ \{taejoonkim, mhashemi\}}@ku.edu
}}
\maketitle

\begin{abstract}

In this paper, we study the distributed computational capabilities of device-to-device (D2D) networks. A key characteristic of D2D networks is that their topologies are reconfigurable to cope with network demands. For distributed computing, resource management is challenging due to limited network and communication resources, leading to inter-channel interference. To overcome this, recent research has addressed the problems of wireless scheduling, subchannel allocation, power allocation, and multiple-input multiple-output (MIMO) signal design, but has not considered them jointly.
In this paper, unlike previous mobile edge computing (MEC) approaches, we propose a joint optimization of wireless MIMO signal design and network resource allocation to maximize energy efficiency. Given that the resulting problem is a non-convex mixed integer program (MIP) which is prohibitive to solve at scale, we decompose its solution into two parts: (i) a resource allocation subproblem, which optimizes the link selection and subchannel allocations, and (ii) MIMO signal design subproblem, which optimizes the transmit beamformer, transmit power, and receive combiner. Simulation results using wireless edge topologies show that our method yields substantial improvements in energy efficiency compared with cases of no offloading and partially optimized methods and that the efficiency scales well with the size of the network.
\end{abstract}


\section{Introduction}
\label{sec:intro}

The number of wireless devices is now over $8.6$ billion, and with the advent of new 5G-and-beyond technologies it is expected to grow to $12.3$ billion by 2022 \cite{chiang2016fog}. 
Many of these devices will be data-processing-capable nodes that facilitate the rapidly growing data-intensive applications running at the network edge, e.g., social networking, video streaming and distributed data analytics. While some devices are occupied processing computationally-intensive applications, e.g., facial recognition, location-based augmented/virtual reality (AR/VR), and online 3D gaming~\cite{yao2019edgeflow, yang2018communication, 8319323}, it may be desirable 
for them 
to offload their data to devices with underutilized resources.
Traditionally, cloud computing architectures have been adopted for such data intensive applications, e.g., Amazon Web Services and Microsoft Azure, but the exponential rise in data generation at the edge is making centralized architectures infeasible for providing latency-sensitive quality of service at scale \cite{chiang2016fog}.

\begin{figure}
\centering
\begin{subfigure}{.48\columnwidth}
  \centering
  \includegraphics[width=\linewidth]{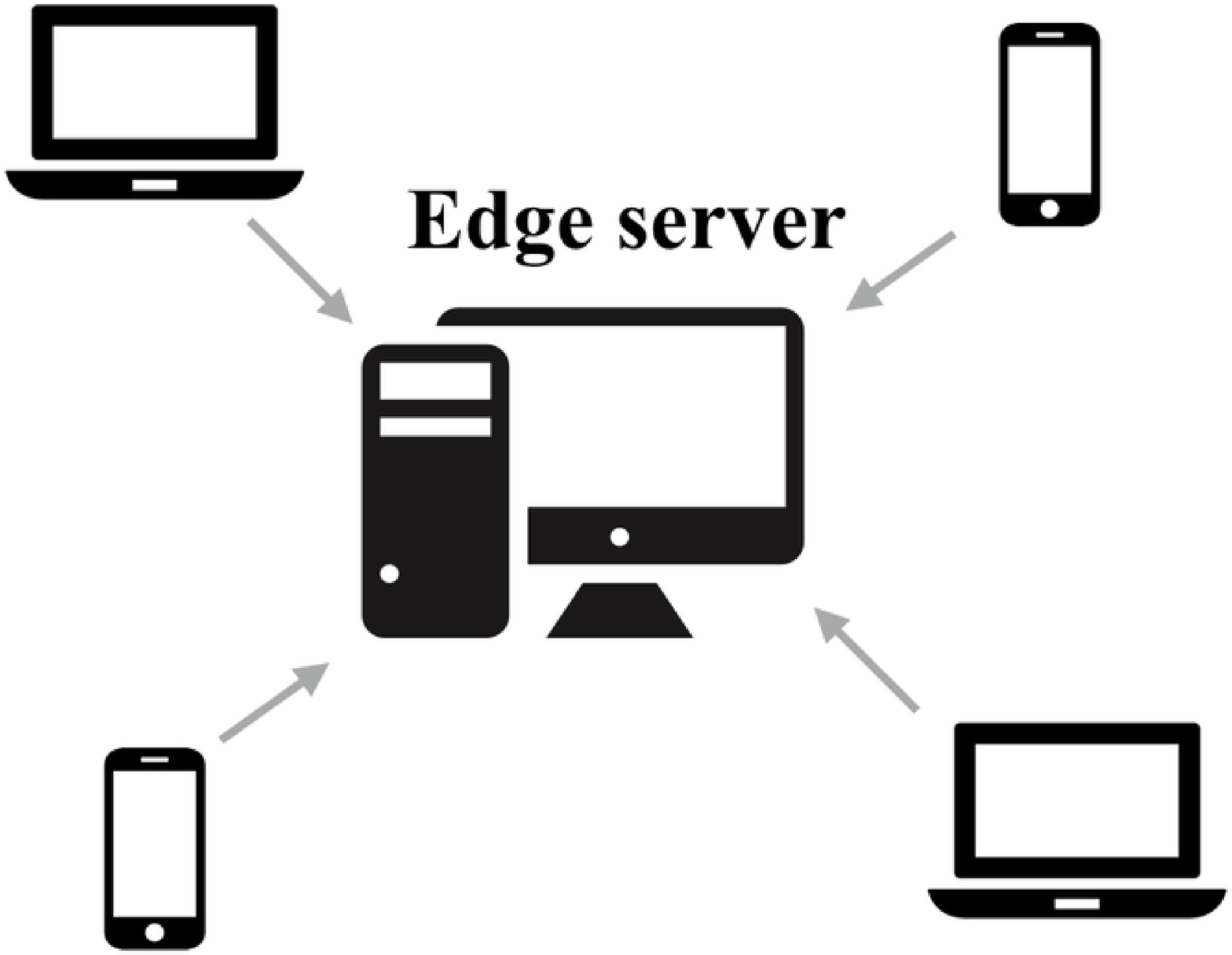}
  \caption{MEC}
  \label{fig:topo:MEC}
\end{subfigure}
\begin{subfigure}{.48\columnwidth}
  \centering
  \includegraphics[width=\linewidth]{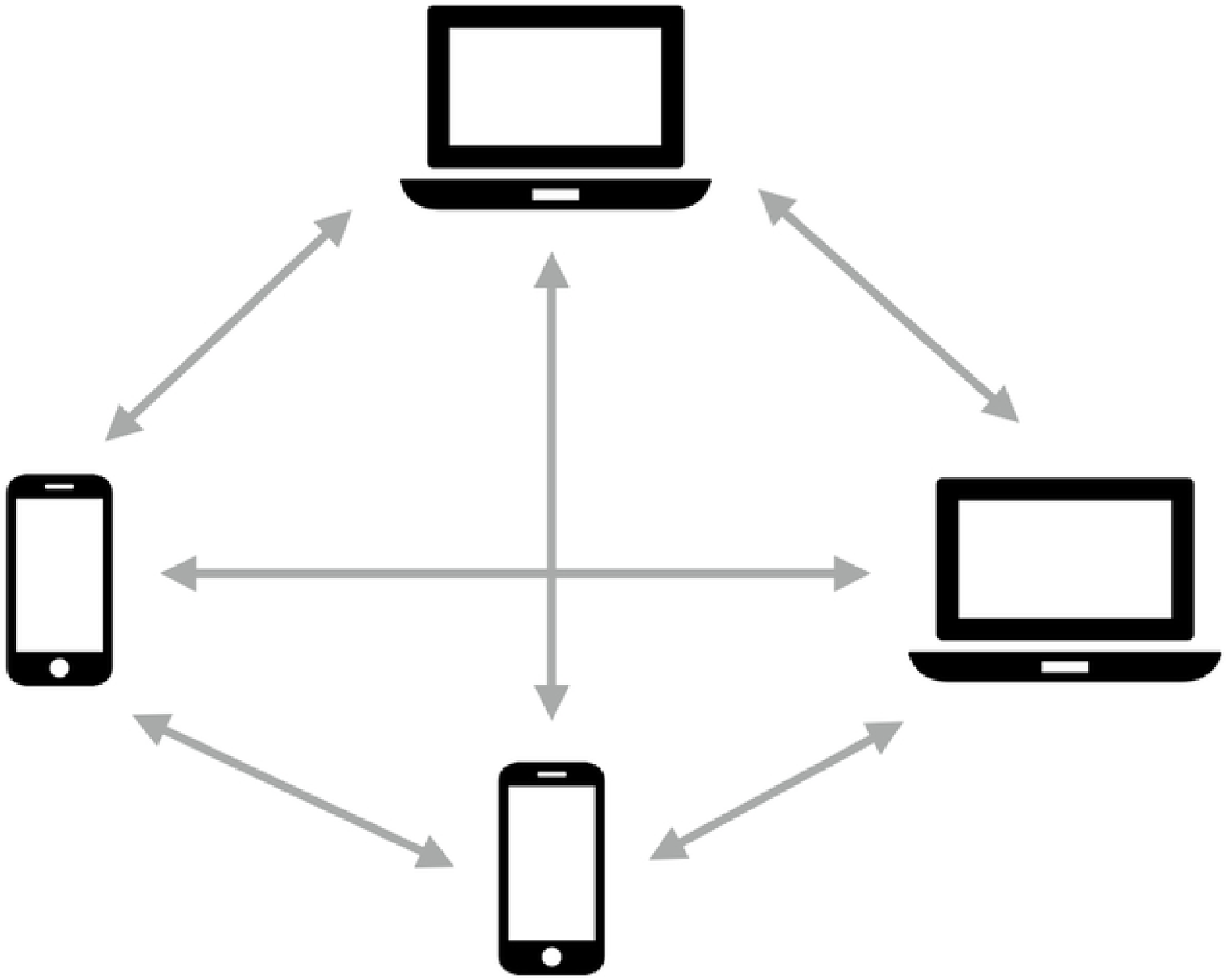}
  \caption{D2D}
  \label{fig:topo:D2D}
\end{subfigure}
\caption{Comparison between the topologies of (a) mobile edge computing (MEC) systems and (b) device-to-device (D2D) networks. MEC topology is typically fixed and predetermined, while D2D topology is not.}
\label{fig:topo}
\vspace{-0.1in}
\end{figure}

As a current trend in wireless networks is reducing cell sizes \cite{Sultan18}, many 5G networks will be dense with short distances, forming a magnitude of smaller subnets \cite{Cisco19}. 
Networks of small subnets combined with improved computational and storage capabilities of edge devices have enabled mobile edge computing (MEC), a recently popularized line of research.
At a high level, MEC leverages radio access networks (RANs) to boost computing power in close proximity to end-users, thus enabling the users to offload their computations to an edge server (central processor)~\cite{you2016multiuser, le2017efficient, messaoudi2017using,  li2019incentive, chen2018decentralized} as shown in Figure \ref{fig:topo}(\subref{fig:topo:MEC}). 
In an MEC architecture, the edge servers 
have high-performance computing units which can process large amounts of computationally intensive tasks efficiently. 
This MEC concept has also been extended to a helper-edge server architecture to exploit the computation resources of idle devices \cite{cao2018joint,xing2019joint}.

The current trend in distributed computing, though, is a migration to even more decentralized architectures.
This is due to the fact that all edge devices can take part in data offloading, given the advances in 5G communication technologies in conjunction with improved computational capabilities of individual devices.
For this reason, device-to-device (D2D) network architectures (shown in Figure \ref{fig:topo}(\subref{fig:topo:D2D})) that were previously studied in 4G LTE standards 
now hold the promise of
providing distributed computing at scale.   

Unlike the MEC system in Figure \ref{fig:topo}(\subref{fig:topo:MEC}), distributed computing in the D2D network of Figure \ref{fig:topo}(\subref{fig:topo:D2D}) will have more complicated topology  management needs given the additional coordination requirements.
In particular, judicious management of network and communication resources is essential for D2D network efficiency because wireless transmissions among edge nodes participating in data offloading will incur inevitable inter-channel interference.
The focus of this paper is on developing such resource management methodologies that jointly optimize network and communication resources 
in D2D networks to adapt to edge computing demands.

\subsection{Related work}
Several existing works have considered resource allocation management in MEC systems. In particular, the problems of wireless scheduling~\cite{molina2014joint,wang2017computation}, interference management through allocation of different subchannels~\cite{wang2017joint}, power allocation~\cite{cao2018joint,wang2017joint2,xing2019joint}, and multiple-input multiple-output (MIMO) signal design~\cite{barbarossa2017enabling, sardellitti2015joint, nguyen2019computation} have each been studied in this context. All of these works, however, have considered the network and communication resources separately, typically focusing on one or the other independently. In reality, these all are interdependent, and there is a need to understand how MEC systems should be designed to jointly optimize link selection and resource management for contemporary edge computing scenarios.
Given the single direction offloading to the edge server,
MEC systems can be viewed as special cases of the D2D networks that we study in this paper (see Figure \ref{fig:topo}).

In terms of D2D networks, some research on D2D networks has addressed resource management problems such as subchannel allocation \cite{han2012subchannel}, power allocation \cite{wen2013energy}, or both of these jointly \cite{wang2013joint,zhao2017joint,6858049}.
Also, MIMO signal design has been considered in scenarios of D2D communications underlaying cellular networks, in order to mitigate inter-channel interferences~\cite{lin2015interplay,tang2013cooperative} or to maximize utility models for cellular users communicating in such scenarios~\cite{wei2013device}.
However, for the above works, the link selection problem is not considered, which is important for network efficiency maximization.
Although some recent works addressed link selection \cite{pu2016d2d} together with device power allocation \cite{liu2018d2d} in D2D networks, 
they in turn have not addressed wireless signal design and subchannel allocation problems.

\subsection{Design Principles and Contributions}
We develop an optimization methodology for network and communication resource management in wireless D2D networks, considering the problems of link selection, subchannel allocation, power allocation, and MIMO signal design jointly. Our approach is driven by two design principles:

\parab{(1) Link selection for energy minimization.} 
The network topology should be defined so that energy efficiency is maximized. 
Given a reconfigurable network topology,
link selection design between nodes in D2D network is more complicated than offloading decisions between nodes and a server 
in MEC systems.

\parab{(2) Joint resource allocation and MIMO signal design.}
The link selection is accompanied with efficient management of available resources.
We jointly consider network resources that involve topology configuration, through link selection and frequency assignment, and communication resources that involve transmission power and MIMO antennas.
Especially, with a large number of links and limited subchannels, MIMO signal design, i.e., beamforming, is essential to mitigate inevitable inter-channel interferences for robust data transfer and optimization.

The contributions of this paper are as follows:
\begin{itemize}
\item We formulate a novel optimization problem for D2D edge computing networks that minimizes the total energy consumption required to process given data samples at the D2D edge nodes, based on a framework for joint wireless MIMO signal design, resource allocation (subchannel and power allocation), and link selection (Section \ref{sec:formulation}).

\item We decompose the integrated framework into two subproblems to make it computationally tractable: one for network resource allocation and the other for MIMO signal design. This facilitates the network resource optimization with integer variables and MIMO signal design optimization with non-integer variables (Section \ref{sec:formulation}).

\item In the optimization of network resource allocation, we jointly determine the link with subchannel allocation. For solving this problem, we modify our objective function and exploit an integer programming. We propose a greedy algorithm as another solution to reduce the computational complexity (Section \ref{sec:algorithm}).

\item With respect to MIMO signal design optimization, we determine the transmit beamformer, transmit power and receive combiner by exploiting interference coordination techniques for MIMO systems. (Section \ref{sec:algorithm}).
The energy minimization problem is converted to the weighted minimum mean square error (WMMSE) minimization problem for tractability (Section \ref{sec:algorithm}).

\item Extensive simulations are presented to evaluate the integrated framework and our proposed algorithms (Section \ref{sec:eval}). 
\end{itemize}

\section{System Model and Optimization Formulation}
\label{sec:formulation}

In this section, we formulate the problem of jointly optimizing MIMO signal design and network resource allocation to minimize the total energy consumption in device-to-device (D2D) network. We first outline preliminaries of the underlying wireless D2D network model in Sec. \ref{ssec:wnm} and then present the optimization in Sec. \ref{ssec:opt}.

\begin{figure}[t]
    \centering
    \includegraphics[width=.8\linewidth]{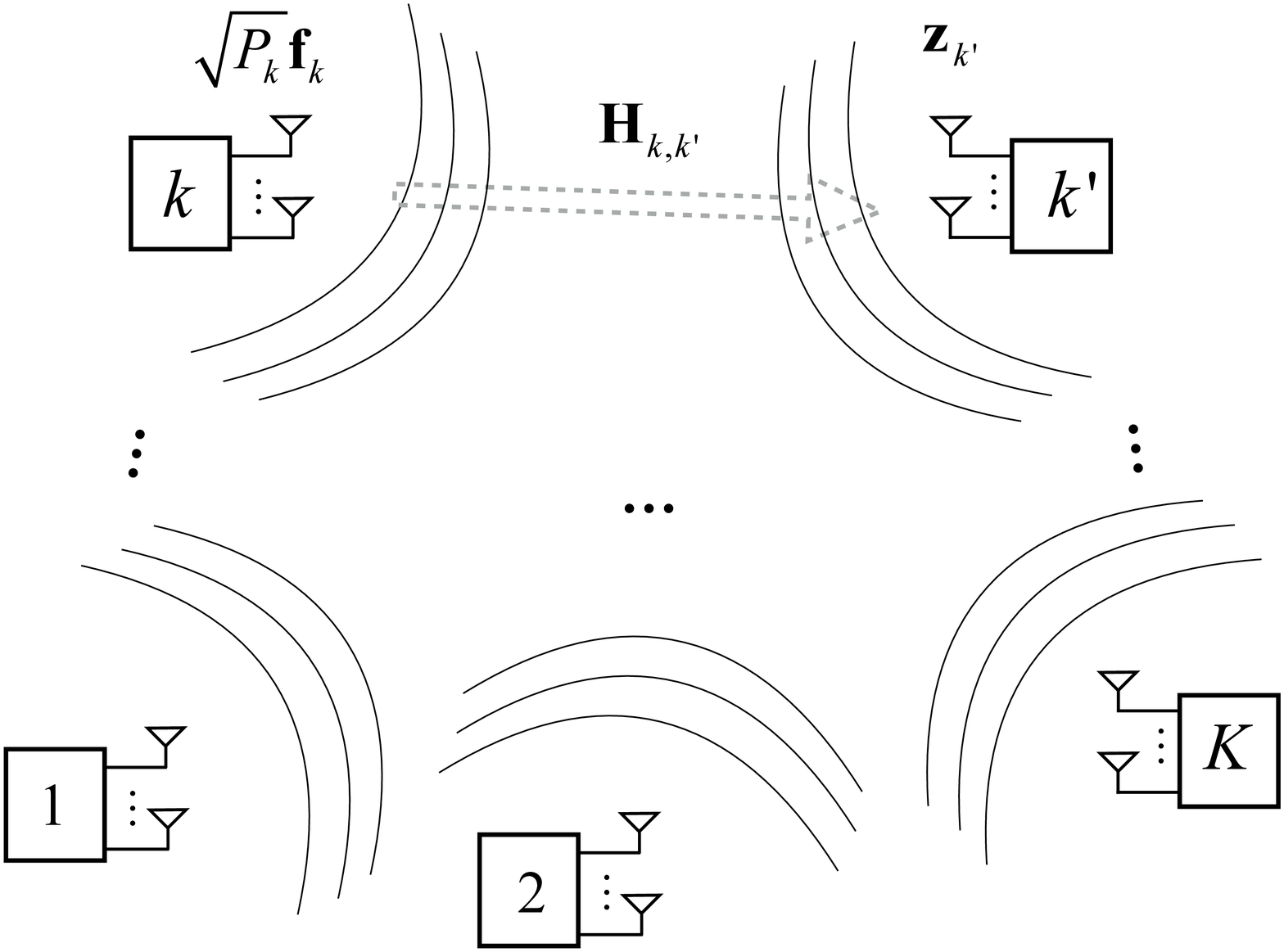}
    \caption{Wireless device-to-device (D2D) network model of a group of $K$ nodes using multiple antennas. 
    }
\label{fig:model:link}
\end{figure}

\vspace{-1mm}
\subsection{Wireless Device-to-device (D2D) Network Model}
\label{ssec:wnm}
Figure 2 demonstrates a wireless D2D network model of $K$ nodes that can transmit using multiple antennas.

\parab{Data transfer.} 
Consider a multi-point to multi-point wireless channel where a group of $K$ nodes can simultaneously transfer (offload) data to one another. With $k,k' = 1,...,K$, we define $a_{k,k'} \in \{0, 1\}$ as the binary variable describing whether node $k$ transmits data to node $k'$, and $I_k$ (in bits) as the length of data present at node $k$ for potential transfer, which is a task eligible for distributed processing. That is, $a_{k,k'} = 1$ implies that the node $k$ sends the data $I_k$ to the node $k'$ while $a_{k,k'} = 0$ means the node $k$ is holding it for local processing. 
Note that data transfer is not allowed from a node to itself, i.e., $a_{k,k} = 0$ for all $k$.

We assume that each task should be processed as a whole, i.e., the task is divisible.
In other words,
If $a_{k,j} = 1$, then $a_{k,k'} = 0$, since the node $k$ will not transmit its data to two or more nodes. We denote the set of link pairs attempting to transmit data as
\begin{equation}
    {\mathcal{A}} = \{ (k,k'): a_{k,k'} = 1\},
\end{equation}
with $(k,k')_n = (k_n,k'_n)$ denoting the $n$-th pair of transmit and receive node where $n=1, ..., L$, and $L = |\mathcal{A}|$ being the total number of link pairs.
For example, if we have the connected link from node 3 to node 1 and it is the first link pair $(n=1)$, then $(k_1,k'_1) = (3,1)$.

\parab{Frequency subchannels.} To facilitate multiple access transmission, the pairs in $\mathcal{A}$ will leverage multiple frequency subchannels $\mathcal{B} = \{b_1, ..., b_{S} \}$, where $S$ is the total number of subchannels, each having the same bandwidth $W$. In general, the subchannels will be limited, i.e., $S \ll L$, meaning subchannels will be shared across multiple links. With ${\bf{m}}_i  \in \{0, 1\}^{L}$ as the binary vector of which active links are using the $i$-th subchannel, $i=1, ..., S$, the $n$-th element $m_{n,i} \in \{0, 1\}$ for $n = 1,..., L$ denotes whether the $n$-th link uses subchannel $i$ ($m_{n,i} = 1$) or not ($m_{n,i} = 0$). 
For example, suppose we have three links and two subchannels. Then $L = 3$ and $S = 2$. One possible subchannel allocation is ${\bf{m}}_1 = (1, 0, 0)^T$ and ${\bf{m}}_2 = (0, 1, 1)^T$, meaning the first link is using the first subchannel $b_1$ and the second and third links are using the second subchannel $b_2$. We also denote $M = [M_{n,m}] \in \{0, 1\}^{L \times L}$ as the binary subchannel co-occurrence matrix $M = \sum_{i=1}^{S} \mathbf{m}_i \mathbf{m}_i^T$; if $M_{n,m} = \sum_i m_{n,i} m_{m,i} = 1$, then links $n$ and $m$ share the same subchannel, and they interfere with each other, and if $M_{n,m} = 0$ they do not.

\parab{Communication demand.} 
The time it takes for the data transfer from node $k$ to $k'$ can be quantified as $I_{k} / R_{k,k'}$, where $R_{k,k'}$ is the maximum achievable data rate over the wireless channel. If a node $k$ has transmit power $P_{k}$, then we can calculate the total energy consumed for data transfer over the set of link pairs $\mathcal{A}$ as in \eqref{eq:em}.
Note that communication only occurs where the link pairs are connected $a_{k,k'} = 1$, , i.e., $(k_n,k'_n)$ pairs for $n=1, ..., L$.
This means that we can write the communication energy $E_M$ as
\begin{align}
\label{eq:em}
E_M = \sum_{(k,k') \in \mathcal{A}} P_{k} \cdot \frac{ I_{k} }{R_{k,k'}}  
= \sum_{n=1}^L P_{k_n} \cdot \frac{ I_{k_n} }{R_{k_n,k'_n}}.
\end{align}
A lower $E_M$ means the device has a lower energy requirement. 
Assuming each node $k$ has $N_{k}^{\text{Tx}}$ transmit antennas and $N_{k}^{\text{Rx}}$ receive antennas,
we consider the case of multiple-input multiple-output (MIMO) links between nodes. The channel matrix from transmitter $k_n$ to receiver $k'_n$, then, is denoted ${\bf{H}}_{k_n,k'_n} \in \mathbb{C}^{N_{k'_n}^{\text{Rx}} \times N_{k_n}^{\text{Tx}}}$. Transmit beamforming at $k_n$ is accomplished by constructing the beamformer $\sqrt{P_{k_n}} {\bf{f}}_{k_n} \in \mathbb{C}^{N_{k_n}^{\text{Tx}}}$ where $|| {\bf{f}}_{k_n} ||_2 = 1$. The receiver $k'_n$ behaves similarly, constructing the combiner ${\bf{z}}_{k'_n} \in \mathbb{C}^{N_{k'_n}^{\text{Rx}}}$ with $||{{\bf{z}}_{{k'_n}}}|{|_2} = 1$. Using these assumptions, the maximum data rate on link $n$ \cite{cover2012} is
\begin{equation}
\label{eq:rate}
    R_{k_n,k'_n} = W \log_2 \left( 1 + \frac{P_{k_n} | {\bf{z}}_{k'_n}^H {\bf{H}}_{k_n,k'_n} {\bf{f}}_{k_n} |^2 }{ \sum\limits_{m \neq n}^{L} M_{n,m} P_{k_m} | {\bf{z}}_{k'_n}^H {\bf{H}}_{k_m,k'_n} {\bf{f}}_{k_m} |^2 + \sigma^2} \right)
\end{equation}
where $W$ is the bandwidth of frequency subchannel, a superscript $H$ denotes the conjugate transpose, and the noise power $\sigma^2 = 1$.
Note that the data rate in (\ref{eq:rate}) considers interference from other links $m \neq n$ sharing the same subchannel allocation.

\parab{Processing demand.} 
There are two possibilities for processing the data $I_k$: the local processing and offloaded processing.
The local processing means that the node $k$ processes its own data $I_k$, i.e., $a_{k,k'} = 0$ for all $k'$.
However, the offloaded processing means that the node $k$ transfers the data $I_k$ to another node $k'$ which then processes it, i.e., $a_{k,k'} = 1$ for some $k'$.
We denote the processing power of node $k$ as $F_k$, which is practically a function of CPU usage and energy efficiency of device \cite{wen2012energy}.
Using this, the energy requirement in each of these cases is
\begin{equation}
\label{eq:ekk}
E_k = F_k \cdot \frac{ I_k }{C_k}, \qquad E_{k,k'} = P_k \cdot \frac{I_k}{R_{k,k'}} + F_{k'} \cdot \frac{I_{k}}{C_{k'}},
\end{equation}
where $C_k$ is the computation speed of node $k$ and $E_{k,k'}$ includes the energy of communication from the node $k$ to $k'$. With this, we can calculate the total energy consumed for data processing in an entire network over $K$ nodes as
\begin{multline}
\label{eq:ep}
    E_P = \sum_{k=1}^{K} \Bigg( \sum_{k' \neq k}^K a_{k,k'} \cdot \Big( P_k \cdot \frac{I_k}{R_{k,k'}} + F_{k'} \cdot \frac{I_{k}}{C_{k'}} \Big) \\
	+ \Big( 1 - \sum_{k' \neq k}^K a_{k,k'} \Big) \cdot F_k \cdot \frac{I_k}{C_k} \Bigg).
\end{multline}

\subsection{Example for Wireless D2D Network Model}


Ideally, the total energy consumption with data offloading in wireless D2D network should be minimized.
To illustrate this, consider a three node example where at most one link can be connected. 
For $k=1, 2, 3$, each node is assumed to have the same length of data $I_k = 10 $ Mbits, computing power $F_k =1$ Watt, transmission power $P_k = 1$ Watt. The transmission data rate is assumed as $R_{k,k'}=2$ Mbits/sec for $k \ne k'$. However, we consider different computing speeds, $C_1 = 10, C_2 = 2$, and $C_3 = 1$ Mbits/sec. Using these definitions, the total energy consumption for local computation should be $E_{P,local} = F_1(I_1/C_1) + F_2(I_2/C_2) + F_3(I_3/C_3) = 16$ Joules. 

The objective is to minimize total energy to find best link pair among
6 possible links: $(1,2)$, $(1,3)$, $(2,1)$, $(2,3)$, $(3,1)$, and $(3,2)$.
It is noteworthy that the link must be properly selected to minimize the total energy.
Suppose we have link $(3,1)$. Then the total energy including transmission energy should be $ E_{P,31} = F_1(I_1/C_1) + F_2(I_2/C_2) + (P_3(I_3/R_{3,1})+F_1(I_3/C_1)) = 12$ Joules, which is less than the local computation energy $E_{P,local}$. This implies that if transmission channel is available and energy-efficient, we can exploit the communication to minimize total energy for data processing.
If we consider link $(2,1)$ instead of link $(3,1)$,
the total energy would be $E_{P,21} = F_1(I_1/C_1) + (P_2(I_2/R_{2,1})+F_1(I_2/C_1)) + F_3(I_3/C_3) = 17$ Joules, which is larger than local computation energy $E_{P,local}$.
Further, we are aware that link $(3,1)$ is the best link pair for minimizing total energy consumption with the relationship $E_{P,31} < E_{P,local} < E_{P,kk'}$ for any $(k,k') \ne (3,1)$, where $E_{P,kk'}$ is the total energy consumption with the link pair $(k,k')$.
Therefore, the link selection optimization is essential for minimizing total energy consumption.
Although this example assumed only one link and identical data rates of the nodes, we are aware that minimizing the inter-channel interferences in many links environment are critical to data rate maximization for selected links, which will lead to total energy minimization.

\subsection{Optimization and Decomposition}
\label{ssec:opt}
We are interested in the combination of MIMO signal design parameters -- i.e., transmit powers $\{P_k\}$, transmit beamformers $\{{\bf{f}}_{k}\}$, and receive combiners $\{{\bf{z}}_{k'}\}$ -- and network resource allocation parameters -- i.e., transmission link pairs $\{a_{k,k'}\}$ and subchannel allocations $\{m_{n,i}\}$ -- that will lead to processing the network data $I_1, ..., I_K$ with minimum energy consumption. 
We can formulate the following optimization problem:
\begin{align}
& \text{minimize}
& & E_P 
\label{eq:obj:EP} \\
& \text{subject to}
& & R_{k,k'} \; \mbox{defined in (\ref{eq:rate})} \;\; \forall a_{k,k'} = 1, \label{eq:con:R} \\
& & & || {\bf{f}}_{k} ||_2 = 1 \;\; \forall k, \;\; || {\bf{z}}_{k'} ||_2 = 1 \;\; \forall k', \label{eq:con:fz} \\
& & & \sum\nolimits_{k=1}^K P_{k} \leq P, \label{eq:con:P} \\
& & & 0 \leq \sum\nolimits_{j=1}^K a_{k,j} + \sum\nolimits_{j=1}^K a_{j,k} \leq 1 \;\; \forall k, \label{eq:con:a1} \\
& & & a_{k,k'} \in \{0,1\} \;\; \forall k,k', \;\; a_{k,k} =0 \;\; \forall k, \label{eq:con:a2} \\
& & & \sum\nolimits_{(k,k') \notin \mathcal{X}} a_{k,k'} = 0, \label{eq:con:a3} \\
& & & \sum\nolimits_{i=1}^S m_{n,i}  = 1 \; \forall n, \; m_{n,i} \in \{0,1\} \; \forall n,i \label{eq:con:m} \\
& \text{variables}
& & \{{\bf{f}}_{k}\}, \{{\bf{z}}_{k'}\}, \{P_{k}\}, \{a_{k,k'}\}, \{m_{n,i}\} \nonumber
\end{align}
Constraints (\ref{eq:con:R}-\ref{eq:con:P}) relate to the MIMO signal design variables, as described in Sec. \ref{ssec:wnm}. 
In \eqref{eq:con:R}, since $R_{k,k'}$ is only defined when $a_{k,k'} =1$, 
the equation \eqref{eq:con:R} can be considered as $R_{k_n,k'_n}$ for $n=1, ..., L$ as in \eqref{eq:em}.
%
In (\ref{eq:con:P}), we impose an overall transmission power budget $P$ that the network cannot exceed. Constraints (\ref{eq:con:a1}-\ref{eq:con:m}), by contrast, restrict the link pair and subchannel allocations, as follows:

\parab{Transmit and receive streams (\ref{eq:con:a1}\&\ref{eq:con:a2}).}
Because each task is indivisible, we consider that each node can have at most one transmit stream, i.e., $\sum\nolimits_{j} a_{k,j} \le 1$.
Also, we assume that each receive node can have at most one receive stream to reduce the complexity of designing the combiner of receive node, i.e., $\sum\nolimits_{j} a_{j,k} \le 1$.
Further, we restrict that each node cannot be in transmit and receive mode simultaneously, which leads to Constraint \eqref{eq:con:a1}.
Constraint \eqref{eq:con:a2} means data transfer is taken as a binary variable and not allowed for same node.


\parab{Link pair candidates (\ref{eq:con:a3}).} Only the link pairs that would lead to a lower energy requirement should be considered for the network through optimization. 
However, we can define the link pair candidates based on comparison of the computation energy, i.e., $F_k \cdot \frac{ I_k }{C_k}$ and $F_{k'} \cdot \frac{ I_k }{C_{k'}}$ for $k \ne k'$, denoting the computation energy for local processing and offloaded processing, respectively.
In other words, if $F_k \cdot \frac{ I_k }{C_k} > F_{k'} \cdot \frac{ I_k }{C_{k'}}$, the pair $(k,k')$ can be the candidate for link pairs.
Therefore, we define the link pair candidate set as $\mathcal{X} = \{(k,k'): F_k \cdot \frac{ I_k }{C_k} > F_{k'} \cdot \frac{ I_k }{C_{k'}}\}$. Note that the optimized link pairs $\mathcal{A}$ should be among the candidates, i.e., $\mathcal{A} \subset \mathcal{X}$.

\parab{Subchannel allocation (\ref{eq:con:m}).} Each link pair $(k_n,k'_n) \in \mathcal{A}$ must be allocated to a subchannel. With the allocation variables $m_{n,i}$ restricted to binary values, the sum over subchannels $i$ for each link $n$ must be $1$.

Assuming all nodes have $N$ transmit and receive antennas, i.e., $N = N_{k}^{\text{Tx}} = N_{k}^{\text{Rx}}$ for all $k$,
the optimization (\ref{eq:obj:EP}-\ref{eq:con:m}) is a mixed integer program (MIP) with $(2N+1)K$ non-integer and $K(K - 1)(1 + S)$ integer variables. 
Also, since the term $P_k \cdot I_k / R_{k,k'}$ as a function of $P_k$ has the form $x / \log(x)$, which is non-convex, the problem is a non-convex MIP. Existing solvers for non-convex MIPs do not scale well with the number of variables \cite{burer2012non}, and even with $K = 10$ nodes equipped with $N = 5$ antennas and $S = 10$ subchannels, our problem already has more than 1000 variables. One of the challenges we address in Sec. \ref{sec:algorithm} is developing an effective algorithm to solve this problem. Shown in Figure \ref{fig:formulation:optimization}, as a first step, we decompose it into two interrelated subproblems, one for the network resource allocation variables (integer) and the other for the MIMO signal design variables (non-integer) as follows:

\begin{figure}[t]
    \includegraphics[trim={0 0 0 75},clip, width=.8\linewidth]{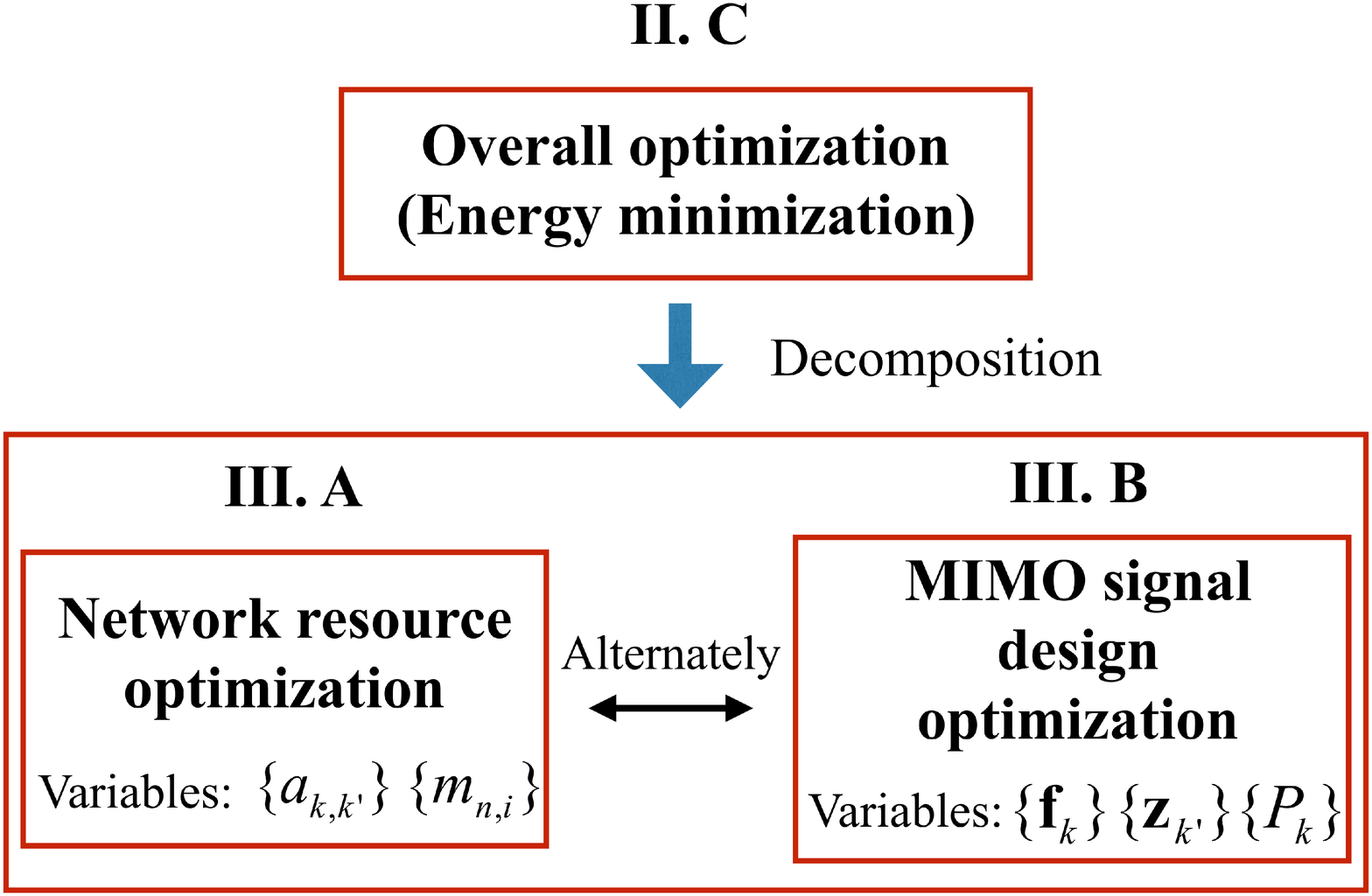}
    \centering
    \caption{Overall optimization is decomposed into two interrelated subproblems: 1) network resource optimization and 2) MIMO signal design optimization.
    }
\label{fig:formulation:optimization}
\end{figure}


\subsubsection{Network resource optimization}
\label{sssec:netopt}
The variables $\{a_{k,k'}\}$ and $\{m_{n,i}\}$ appear in each term of the objective (\ref{eq:obj:EP}). With $\{{\bf{f}}_{k}\}$, $\{{\bf{z}}_{k'}\}$, and $\{P_{k}\}$ fixed, then, we have the following optimization:
\begin{align}
& \text{minimize}
& & E_P
\nonumber \\
& \text{subject to}
& & \mbox{Constraint } (\ref{eq:con:R}), \; (\ref{eq:con:a1}-\ref{eq:con:m}) \nonumber \\
& \text{variables}
& & \{a_{k,k'}\}, \{m_{n,i}\} \nonumber
\end{align}
Given the dependence of $R_{k,k'}$ on these variables, this problem is a non-linear integer program, but is no longer an MIP.


\subsubsection{MIMO signal design optimization}
\label{sssec:sigopt}
The variables $\{{\bf{f}}_{k}\}$, $\{{\bf{z}}_{k'}\}$, and $\{P_{k}\}$ only appear in the term of (\ref{eq:obj:EP}) captured by the communication demand $E_M$ in (\ref{eq:em}). With $\{a_{k,k'}\}$ and $\{m_{n,i}\}$ fixed, it reduces to the following optimization:
\begin{align}
& \text{minimize}
& & E_M
\nonumber \\
& \text{subject to}
& & \mbox{Constraints } (\ref{eq:con:R}-\ref{eq:con:P}) \nonumber \\
& \text{variables}
& & \{{\bf{f}}_{k}\}, \{{\bf{z}}_{k'}\}, \{P_{k}\} \nonumber
\end{align}
Given the dependence of $R_{k,k'}$ on these non-integer variables, the problem is a non-convex optimization, but is also no longer an MIP. In fact, the dependence of $R_{k,k'}$ on both sets of variables renders the optimization (\ref{eq:obj:EP}-\ref{eq:con:m}) non-separable, and hence the subproblems must be solved alternately \cite{chiang2007layering}. 
Although the decomposition for the overall optimization problem does not enable to obtain optimal solution, the obtained (suboptimal) solution from the decomposition can reduce the energy consumption significantly by efficient data offloading compared to that of local processing.
This is approximately a $17\%$ reduction in the Figure \ref{fig:link} example.
We will develop this decomposition approach in Section \ref{sec:algorithm}.

\section{Data Transfer and Processing Algorithms}
\label{sec:algorithm}

\subsection{Network resource optimization}
\label{ssec:nro}
We propose two different algorithms to solve for $\{a_{k,k'}\}$ and $\{m_{n,i}\}$ with $\{{\bf{f}}_{k}\}$, $\{{\bf{z}}_{k'}\}$ and $\{P_{k}\}$ fixed.

\vspace{+2mm}
\subsubsection{Non-linear integer program}
\label{sssec:nlip}
The variables, $\{a_{k,k'}\}$ and $\{m_{n,i}\}$ cannot be separable because the subchannel allocation $m_{n,i}$ is determined only for the connected link $a_{k,k'}=1$. 
This means that, in order to find the best combinations of $\{a_{k,k'}\}$ and $\{m_{n,i}\}$, 
the subchannel should be determined jointly with link pairs.
Therefore, for joint optimization for $\{a_{k,k'}\}$ and $\{m_{n,i}\}$, 
three dimensional (3D) variable, $a_{k,k'}^i$ is introduced, meaning whether $k$ transmits data to $k'$ through the subchannel $i$.
The variable $\{a_{k,k'}^i\}$ is linked with $\{a_{k,k'}\}$ and $\{m_{n,i}\}$ by
\begin{equation}
    a_{k,k'} = \sum\nolimits_{i = 1}^{S} a_{k,k'}^i, \quad
    m_{n,i}=a_{k_n, k'_n}^i,
    \label{eq:akki}
\end{equation}
where the meaning of two relationships is that the link pair $(k,k')$ is mapped to only one subchannel and that $n$-th link pair $(k_n,k'_n)$ is using subchannel $i$.



Then, by substituting $a_{k,k'}^i$ for $a_{k,k'}$ and $m_{n,i}$, the formulation \eqref{eq:ep} can be rewritten as 
\begin{multline}
\label{eq:ep3D}
E_P = \sum_{k=1}^{K} \Bigg( \sum_{k' \neq k}^K \sum_{i=1}^{S} a_{k,k'}^i \cdot \Big( P_k \cdot \frac{I_k}{R_{k,k'}^i} + F_{k'} \cdot \frac{ I_{k'}} {C_{k'}} \Big) \\
	+ \Big( 1 - \sum_{k' \neq k}^K \sum_{i=1}^{S} a_{k,k'}^i \Big) \cdot F_k \cdot \frac{I_k}{C_k} \Bigg),
\end{multline}
where
\begin{equation}
    \label{eq:rate3D}
    R_{k,k'}^i = \log _2 \Bigg( 1 + \frac{{ P_k | {\bf{z}}_{k'}^H {\bf{H}}_{k,k'}{\bf{f}}_k}{|^2}} 
    {\sum\nolimits_{l \ne k, k'}^{K} \sum\nolimits_{l'=1}^K {a_{l,l'}^i {P_l} |{\bf{z}}_{k'}^H {\bf{H}}_{l,k'} {\bf{f}}_l} |^2   + 1} \Bigg).
\end{equation}

The constraints $ (\ref{eq:con:a1} - \ref{eq:con:a3} )$ change to 
\begin{gather}
0 \leq \sum\nolimits_{j=1}^K \sum\nolimits_{i=1}^{S} a_{k,j}^i + \sum\nolimits_{j=1}^K \sum\nolimits_{i=1}^{S} a_{j,k}^i \leq 1 \;\; \forall k, \label{eq:con:a1'} \\
a_{k,k'}^i \in \{0,1\} \;\; \forall k,k',i, \;\;\;\; \sum\nolimits_{i=1}^S a_{k,k}^i =0 \;\; \forall k, \label{eq:con:a2'} \\
\sum\nolimits_{(k,k') \notin \mathcal{X}} \sum\nolimits_{i=1}^{S} a_{k,k'}^i = 0, \label{eq:con:a3'}
\end{gather}
where by introducing $\{a_{k,k'}^i\}$, the constraint ($\ref{eq:con:m}$) is satisfied since we get $\sum_{i=1}^S m_{n,i} = \sum_{i=1}^S a_{k_n,k'_n}^i = a_{k_n,k'_n} = 1$ from the two relationships in \eqref{eq:akki}. Note that $n$-th link pair $({k_n,k'_n})$ comes from the connected link $\mathcal{A}$, which leads to $a_{k_n,k'_n} = 1$.

Then, we obtain a non-linear integer program problem:
\begin{align}
& \text{minimize}
& & E_P \; \mbox{in} \; \eqref{eq:ep3D} \nonumber \\
& \text{subject to}
& & \mbox{Constraints } (\ref{eq:con:a1'}-\ref{eq:con:a3'}) \nonumber \\
& \text{variables}
& & \{a_{k,k'}^i\} \nonumber
\end{align}

We implemented this algorithm using the Solving Constraint Integer Programs (SCIP) solver \cite{scip2009} in the Matlab environment.
Since it is {\it{NP-hard}} problem, a large computational time is required when the number of variables is large. For example, for running with $K=20$ and $S=2$, we have total $K \times K \times S =800$ variables and a computation time of approximately an hour on a 3.5 GHz CPU.
\vspace{+2mm}

\subsubsection{Greedy algorithm}
\label{sssec:greedy}

For all link pairs $(k,k')$ for $k,k'=1, .., K$, we find the best pair to maximize the saved energy by data offloading, $D_{k,k'}$ as
\begin{align}
    \underset{  (k,k') } {\text{maximize}} \;\;\;\;\; D_{k,k'} = E_k - E_{k,k'},
    \label{eq:se}
\end{align}
where $E_k$ and $E_{k,k'}$ are given in \eqref{eq:ekk}. The $(k,k')$ link will be connected when $D_{k,k'} > 0$.
The link pairs are sequentially determined until all possible links are connected, which leads to $a_{k,k'}=1$ when links are connected and $a_{k,k'}=0$, otherwise.

Given the possible link candidates $(k,k') \in \mathcal{X}$, we need
to allocate the subchannels to the link pairs to minimize the interference.
At the beginning, when new subchannels are available,
we allocate the empty subchannel which is interference-free, to the seltected link pair until all subchannels are used.
However, after allocating all interference-free subchannels to the link pairs, the subchannel allocation to the remaining link pairs will incur inter-channel interference.  
In this case, we should compare all combinations of link pairs and subchannels and choose the best combination.
We performed this process until all possible pairs are determined.
During this process, whenever we determine the best pair, we update the candidate set $\mathcal{E}$ with the constraint \eqref{eq:con:a1} i.e., removing the candidates from the determined links so far.
A formal description of the algorithm for joint link pair and subchannel is provided in Algorithm \ref{al:greedy}.
%
 \begin{algorithm}[t]
 \caption{Greedy algorithm for joint link pair and subchannel optimization}
 \label{al:greedy}
 \begin{algorithmic}[1]
 \small
  \STATE \textbf{Initialize} 
  \STATE $a_{k,k'}^i = 0$ $\forall (i,k,k')$, $s = 1$, $\mathcal{E} = \mathcal{X}$ (initial candidate set)
  \STATE ${e_{k,k'}} = 1$ for $(k,k') \in \mathcal{E}$, ${e_{k,k'}} = 0$ otherwise.
  \REPEAT
  \IF{$s \le S$} 
    \STATE \text{1. Allocate new subchannel to the link pair}
    \STATE ${(k_p,k'_p)} = \underset{ (k,k') \in \mathcal{E} } {\text{maximize}} \;\; {D_{k,k'}}$, where 
    \STATE ${D_{k,k'}} = {F_k} {\frac{{I_k}}{{{C_k}}}} - \big( {{P_k}\frac{{I_k}}{{{R_{k,k'}}}} + {F_{k'}}\frac{{I_k}}{{{C_{k'}}}}} \big) $
    \STATE $R_{k,k'} = \log _2 (1 + P_k|{\bf{z}}_{k'}^H {\bf{H}}_{k,k'}{\bf{f}}_k |^2 ) $
    \IF{${D_{{k_p},k{'_p}}} \le 0$} 
    \STATE \textbf{terminate}
    \ENDIF
    \STATE $a_{{k_p},k{'_p}}^s = 1$, \; $s \leftarrow s + 1$
  \ELSE
    \STATE \text{2. Allocate the used subchannel to the link pair}  
    \STATE ${(i_p,k_p,k'_p)} = \underset{ i,(k,k') \in \mathcal{E} } {\text{maximize}} \;\; D_{k,k'}^i $, where 
    \STATE $ D_{k,k'}^i =  {F_k} {\frac{{I_k}}{{{C_k}}}} - \big( {{P_k}\frac{{I_k}}{{R_{k,k'}^i}} + {F_{k'}}\frac{{I_k}}{{{C_{k'}}}}} \big) $ and $R_{k,k'}^i$ in (\ref{eq:rate3D})
    \IF{${D_{i_p,k'_p}^{i_p}} \le 0$}
    \STATE \textbf{terminate}
    \ENDIF
    \STATE $a_{{k_p},k{'_p}}^{{i_p}} = 1$ 
  \ENDIF
    \STATE Update candidate set $\mathcal{E} = \{ (k,k'):{e_{k,k'}} = 1\} $ from Constraint \eqref{eq:con:a1}, $ \sum_{j = 1}^K {e_{k,j}^{}}  + \sum_{j =1}^K {e_{j,k}^{}}  = 0$ for $k = k_p, k'_p$.
  \UNTIL { $|\mathcal{E}| > 0$ }
  \RETURN $a_{k,k'}^i$ $\forall (i,k,k')$
 \end{algorithmic}
 \end{algorithm}

\subsection{MIMO signal design optimization}
\label{ssec:sdo}
We turn to the optimization of the MIMO signal design variables, $\{{\bf{f}}_{k}\}$, $\{{\bf{z}}_{k'}\}$ and $\{P_{k}\}$ with $\{a_{k,k'}\} $ and $\{m_{n,i}\}$ fixed.
Instead of minimizing \eqref{eq:em}, we desire to convert the problem to familiar problem and find the sub-optimal solution for communication demand. We will use the relationship $\sum_i {(x_iy_i)} \le (\sum_i {x_i}) (\sum_i y_i)$ for $x_i,y_i \ge 0 $, and minimize the upper bound for the communication demand in \eqref{eq:em}, $E^{up}_M$, which is given as
\begin{equation}
\label{eq:em'}
E^{up}_M = P \sum\nolimits_{n=1}^L \frac{ I_{k_n} }{R_{k_n,k'_n}},
\end{equation}
where $P$ is the total transmission power budget in (\ref{eq:con:P}). 
The problem in \eqref{eq:em'} is simplified to a problem for minimizing the time delay, which has a similar form as maximizing harmonic-rate \cite{luo2008dynamic}. 
%
Combining the power $P_{k_n}$ and ${{\bf{f}}_{k_n}}$ into ${{\bf{g}}_{k_n}}$, i.e., ${{\bf{g}}_{k_n}} = \sqrt {{P _{k_n}}} {{\bf{f}}_{k_n}} $, leads to
\begin{align}
    & {\text{minimize}} 
    & & T = \sum\nolimits_{n = 1}^{{L}} {\frac{{I_{{k_n}}}}{{{R_{{k_n},k{'_n}}}}}} \label{eq:T} \\
    & {\text{subject to} } & & \sum\nolimits_{n=1}^L || {\bf{g}}_{{k_n}} ||_2^2 \le P, \;\; || {\bf{z}}_{{k'_n}} ||_2 = 1 \label{eq:con:fzP}
    \\
    & {\text{variables} } & & \{{\bf{g}}_{k_n}\}, \{{\bf{z}}_{k'_n}\}
    \label{eq:var:fz}
\end{align}
where ${R_{k_n,k'_n}}$ is given in \eqref{eq:rate} and the constraint \eqref{eq:con:fzP} comes from (\ref{eq:con:fz}) and (\ref{eq:con:P}).
This problem is non-convex and {\it NP-hard}. Therefore, suboptimal solution can be found by alternatively solving for the variables in \eqref{eq:var:fz}.

\subsubsection{Receive combiner}
First, with $\{{{\bf{g}}_{k_n}}\}$ fixed, we will solve for $\{{\bf{z}}_{k'_n}\}$. 
Because with $\{{{\bf{g}}_{k{_n}}}\}$ fixed, each $R_{{k_n},k{'_n}}$ solely depends on ${{\bf{z}}_{k{'_n}}}$,
the optimization problem is decoupled across receivers, resulting in the SINR maximization problem for individual receiver
\begin{align}
    \underset{ { {\bf{z}}_{k'_n}  }} {\text{ maximize } } 
    \frac{| {\bf{z}}_{k'_n}^H {\bf{H}}_{k_n,k'_n} {\bf{g}}_{k_n} |^2}
    {\sum_{m \ne n}^{L} 
    {{M_{n,m}} |{\bf{z}}_{k'_n}^H {{\bf{H}}_{k_m,k'_n}} {{\bf{g}}_{k_m}}| ^2}  + 1}.
    \label{eq:z:sinr}
\end{align}
The solution to \eqref{eq:z:sinr} is given by the minimum mean square error (MMSE) receiver \cite{li2005distribution},
\begin{gather}
    {\bf{z}}_{k{'_n}}^{mmse} = {\bf{J}}_{n}^{-1}( {\bf{g}} ) {{\bf{H}}_{{k_n},k{'_n}}}{{\bf{g}}_{{k_n}}},
    \label{eq:zmmse}
\end{gather}
where 
\begin{equation}
    {\bf{J}}_{n}( {\bf{g}} ) = \sum\nolimits_{m = 1}^{{L}} {{M_{n,m}}{{\bf{H}}_{{k_m},k{'_n}}}{{\bf{g}}_{{k_m}}}}  {\bf{g}}_{{k_m}}^H {\bf{H}}_{{k_m},k{'_n}}^H + {\bf{I}}.
    \label{eq:J}
\end{equation}
Here, ${\bf{I}}$ denotes the $N_{k'_n}^{\text{Rx}} \times N_{k'_n}^{\text{Rx}}$ identity matrix.
\vspace{+2mm}

\subsubsection{Transmit beamformer and power}
Next, fixing $\{{{\bf{z}}_{k'_n}}\}$ in \eqref{eq:zmmse}, we solve for \eqref{eq:T} to design $\{{{\bf{g}}_{k_n}}\}$. 
Unlike solving for $\{{{\bf{z}}_{k'_n}}\}$, $\{{{\bf{g}}_{k_n}}\}$ cannot be decoupled across transmitters since ${R_{k_n,k'_n}}$ depends on other beamformers besides ${{\bf{g}}_{k_n}}$.
This can be handled by transferring the original problem to the weighted minimum mean square error (WMMSE) problem \cite{Qingjiangshi} with ${{\bf{z}}_{k'_n}} = {\bf{z}}_{k{'_n}}^{mmse}$ for all $k'_n$ given by
%
%
\begin{align}
     \underset{ \{{\bf{g}}_{k_n}\}, \{w_n\} } {\text{minimize}} \sum\nolimits_{n = 1}^{{L}} {I_{{k_n}}({w_n}{e_n} + c(\gamma ({w_n})) - {w_n}\gamma ({w_n}))},
    \label{eq:wmmse}
\end{align}
where $w_n$ is auxiliary weight variable, 
$c(\cdot) = -1/\log(\cdot)$ and
$\gamma(y)$ is the inverse function of $y= \frac{\partial }{{\partial x}} c (x)$.
We note that the explicit form of $\gamma(y)$ is not needed to solve for $\{{{\bf{g}}_{k_n}}\}$ in our problem. 
The mean square error at receive node $k'_n$ in $n$-th link pair is then given by
\begin{multline}
    {e_n} = | 1 - {\bf{z}}_{k{'_n}}^H{{\bf{H}}_{{k_n},k{'_n}}}{{\bf{g}}_{{k_n}}} |^2 + 
    \\
    \sum\nolimits_{m \ne n}^{{L}} {M_{n,m}} {\bf{z}}_{k{'_n}}^H {{\bf{H}}_{{k_m},k{'_n}}} {{\bf{g}}_{{k_m}}} {\bf{g}}_{{k_m}}^H {\bf{H}}_{{k_m},k{'_n}}^H  {\bf{z}}_{k{'_n}}  +1.
    \label{eq:mse}
\end{multline}
Note that the problem \eqref{eq:wmmse} is convex for each variable when the others are fixed.
Minimizing \eqref{eq:T} with respect to $\{{\bf{g}}_{{k_n}}\}$ is now accomplished by solving \eqref{eq:wmmse} for $\{{\bf{g}}_{{k_n}}\}$ and $\{w_n\}$. 


With $\{{{\bf{z}}_{k'_n}}\}$ and $\{{{\bf{g}}_{k_n}}\}$ fixed, the derivatives of \eqref{eq:wmmse} with respect to $w_n$ gives the equation $e_n - \gamma ({w_n}) = 0$, of which proof is given in \cite{Qingjiangshi}. Note that the equation $e_n - \gamma ({w_n}) = 0$ is equivalent to $\frac{\partial }{{\partial e_n}}c(e_n) - {w_n} = 0$.
Then, we obtain
\begin{align}
    w_n = \frac{\partial }{{\partial e_n}} c({e_n}) = \frac{1}{{{e_n}{{(\log ({e_n}))}^2}}},
\end{align}
where $e_n$ in \eqref{eq:mse} is given by $\{{{\bf{z}}_{k'_n}}\}$ and $\{{{\bf{g}}_{k_n}}\}$.

We fix the combiner $\{{{\bf{z}}_{k'_n}}\}$ and the auxiliary weight $\{w_n\}$, and optimize the beamformer $\{{{\bf{g}}_{k_n}}\}$ by solving \eqref{eq:wmmse}.
Substituting $\{e_n\}$ found from \eqref{eq:mse} in \eqref{eq:wmmse} makes the beamformers ${{\bf{g}}_{k_n}}$ for all $k_n$ decoupled across the links, resulting in the following optimization problem:
%
%
\begin{align}
    & \underset{  \{{\bf{g}}_{{k_n}}\}  } {\text{minimize}}  \;\;\;
    {\sum\nolimits_{n = 1}^{{L}} {{I_{{k_n}}}{w_n} |1 -  {\bf{z}}_{k{'_n}}^H {{\bf{H}}_{{k_n},k'_n}} {{\bf{g}}_{{k_n}}}|^2 }}
    \label{eq:fconv}
    \\
    &+ \sum\nolimits_{n = 1}^{{L}} \sum\nolimits_{m \ne n}^{{L}} {{I_{{k_m}}}{w_m}{M_{n,m}}{\bf{z}}_{{k'_m}}^H{{\bf{H}}_{{k_n},{k'_m}}}{{\bf{g}}_{{k_n}}} {\bf{g}}_{{k_n}}^H {\bf{H}}_{{k_n},{k'_m}}^H {\bf{z}}_{{k'_m}}}
    \nonumber
    \\
    & \text{subject to} \;\;\; \sum\nolimits_{n = 1}^{L} ||{{\bf{g}}_{{k_n}}}||_2^2 \le P.
    \label{eq:con:f}
\end{align}
This problem is a standard quadratic convex optimization problem, which can be solved by using Karush-Kuhn-Tucker (KKT) conditions \cite{boyd2004convex}. Since it is a standard procedure, we omit the details here. 
An algorithmic description for the MIMO signal design is outlined in Algorithm \ref{al:fz}.


 \begin{algorithm}[t]
 \caption{Algorithm for transmit beamformer containing power and receive combiner optimization}
 \label{al:fz}
 \begin{algorithmic}[1]
 \small
  \STATE \textbf{Initialize} 
  \STATE $w_n = 2, w'_n=1$
  \REPEAT
    \STATE {
    ${\bf{z}}_{k{'_n}}^{mmse} = {\bf{J}}_{n}^{-1} ( {\bf{g}} ) {{\bf{H}}_{{k_n},k{'_n}}}{{\bf{g}}_{{k_n}}}$ $\forall n$ 
    where ${\bf{J}}_{n} ( {\bf{g}} )$ in \eqref{eq:J}
    }
    \STATE $w'_n \leftarrow w_n$  $\forall n$
    \STATE $w_n = 1/\big( {e_n}{{(\log ({e_n}))}^2} \big)$ $\forall n$ where $e_n$ in \eqref{eq:mse}  
    \STATE Find ${{\bf{g}}_{k_n}}$ by solving convex problem \eqref{eq:fconv}, \eqref{eq:con:f}.
  \UNTIL { $ \big| \sum_{n=1}^{L} \log_2 w_n - \sum_{n=1}^{L} \log_2 w'_n \big| \ge \varepsilon$
  }
  \RETURN ${{\bf{f}}_{k_n}} = {{\bf{g}}_{k_n}}/||{{\bf{g}}_{k_n}}||_2$, \; $P_{k_n} = ||{{\bf{g}}_{k_n}}||_2^2$,
  \\ \quad \quad \quad \; and $ {\bf{z}}_{k{'_n}} = {\bf{z}}_{k{'_n}}^{mmse} / ||{\bf{z}}_{k{'_n}}^{mmse}||_2 $ \; $\forall n$
 \end{algorithmic}
 \end{algorithm}

\section{Algorithm Evaluation and Discussion}
\label{sec:eval}

In this section, we perform simulations to validate the proposed framework for MIMO signal design and resource allocation through our proposed algorithm.

\subsection{Simulation setup}
\label{ssec:total}

For resource allocation, we can apply one of our two proposed schemes, the non-linear integer programming (NLIP) in \ref{sssec:nlip} or greedy algorithm in \ref{sssec:greedy}.
For MIMO signal design, the WMMSE algorithm in \ref{ssec:sdo} is used.
Therefore, we evaluate two combinations of algorithm, NLIP--WMMSE and Greedy--WMMSE, which will be denoted by NLIP and Greedy in Figs. \ref{fig:iter}-\ref{fig:node}.
The decomposed optimization depends on  initial values of $\{{\bf{f}}_k\}$, $\{{\bf{z}}_{k'}\}$, $\{P_k\}$, $\{a_{k,k'}\}$, and $\{m_{n,i}\}$. 
For our proposed algorithm, NLIP--WMMSE and Greedy--WMMSE, we run each of them 10 times with different initializations and keep the {\it{mean}} of the results and the minimizer, called the {\it{best}} in Figure \ref{fig:iter}-\ref{fig:node}.
The minimizer is selected as the one minimizing the total energy the most.
For initializations, the link pair $\{a_{k,k'}\}$ is randomly chosen from link cadidate set $\mathcal{X}$, the subchannel $\{m_{n,i}\}$ is allocated to connected links $a_{k,k'}=1$ randomly, $P_k = P/L_{max}$, and $\{{\bf{f}}_k\}$, $\{{\bf{z}}_{k'}\}$ are generated to be uniformly distributed on the complex unit sphere~\cite{au2007performance}, where $L_{max} = \lfloor K/2 \rfloor$ is the maximum number of link pairs with total $K$ nodes.

For all our simulations, we generated random Gaussian channel ${\bf{H}}_{k,k'}$ for $k,k' = 1,...,K$, and used the bandwidth $W=1$ MHz for each subchannel, and $P=5$ Watt with noise power $\sigma^2 = 1$.
As for computational speed $C_k$, the general unit Hz (or cycles/s) can be converted to bits/sec according to the application it processes. 
For example, in the work of the audio signal detection \cite{johnson2006modified}, 500 cycles are required to process 1 bit, i.e., 500 cycles/bit. In other words, 500 MHz CPU can be converted to 1 Mbits/sec.
As an example, for computation power $F_k$ the mobile devices with $500$ MHz CPU require about 0.6 Watt for CPU processing \cite{ardito2013profiling}.
Based on trying to emulate heterogeneous devices with different energy efficiency, we generate the length of data $I_k \sim U(1,20)$ Mbits, computational speed $C_k \sim U(0.1,2)$ Mbits/sec, and computing power $F_k \sim U(0.5,1)$ Watt for $k=1, ..., K$, where $U(a,b)$ is continuous uniform distribution ranging from $a$ to $b$.
All nodes are assumed to have $N$ transmit and receive antennas, i.e., $N = N_{k}^{\text{Tx}} = N_{k}^{\text{Rx}}$ for all $k$.
For all our simulations, we select the number of antennas based on properness of interference alignment \cite{yetis2010}, $2N \ge L_{max} + 1$. 


\subsection{Varying number of links}
\label{ssec:total}

\begin{figure}
    \centering
  \includegraphics[width=0.95\linewidth]{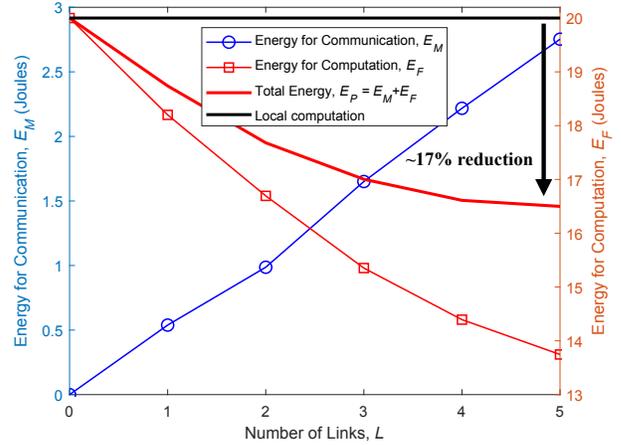}
  \caption{The relationship among communication, computation and total energy across the number of links. The total energy expended decreases substantially as the available communication resources to our optimization are increased.}
  \label{fig:link}
\end{figure}

Figure \ref{fig:link} depicts the relationship among the total energy $E_P$, communication energy $E_M$, and computation energy $E_F$, along the number of links $L$ where $E_F$ can be calculated as $E_F = E_P - E_M$. 
Interestingly, \emph{the total energy $E_P$ decreases below the local computation energy as the available communication resources (links) are more heavily exploited}, while the energy for communication increases and that for computation decreases.
In other words, since more nodes participate in offloading, we can further reduce energy consumption upto approximately 17 $\%$, reducing energy from 20 Joules to 16.6 Joules.
This simulation was performed by using Greedy-WMMSE, when $K=10$ nodes, $N=6$ antennas for both transmit and receive and $S=3$ subchannels.

\subsection{Algorithm convergence analysis}
\label{ssec:iter}

\begin{figure}
\begin{subfigure}{.48\textwidth}
  \centering
  \includegraphics[width=\linewidth]{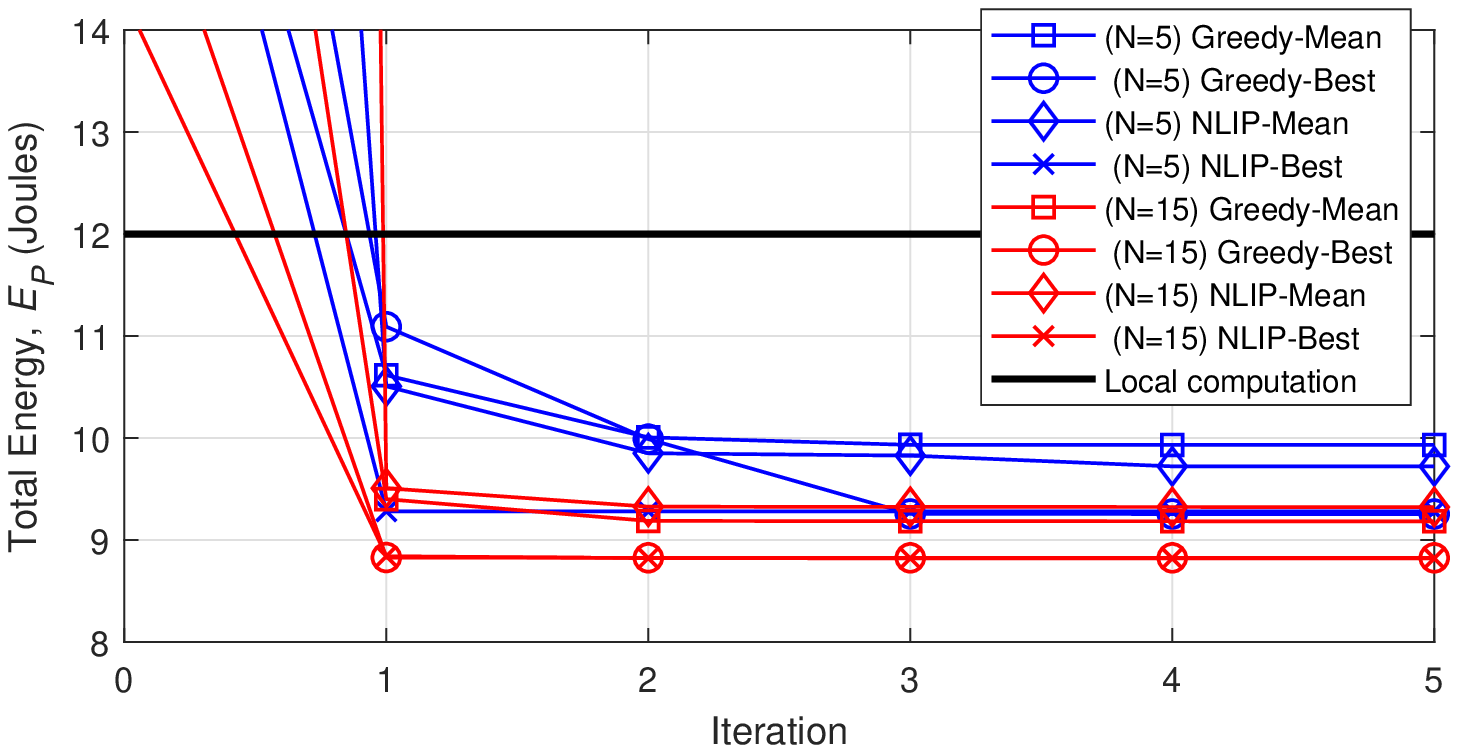}
  \label{fig:iter:EP}
\end{subfigure}
\begin{subfigure}{.48\textwidth}
  \centering
  \includegraphics[width=\linewidth]{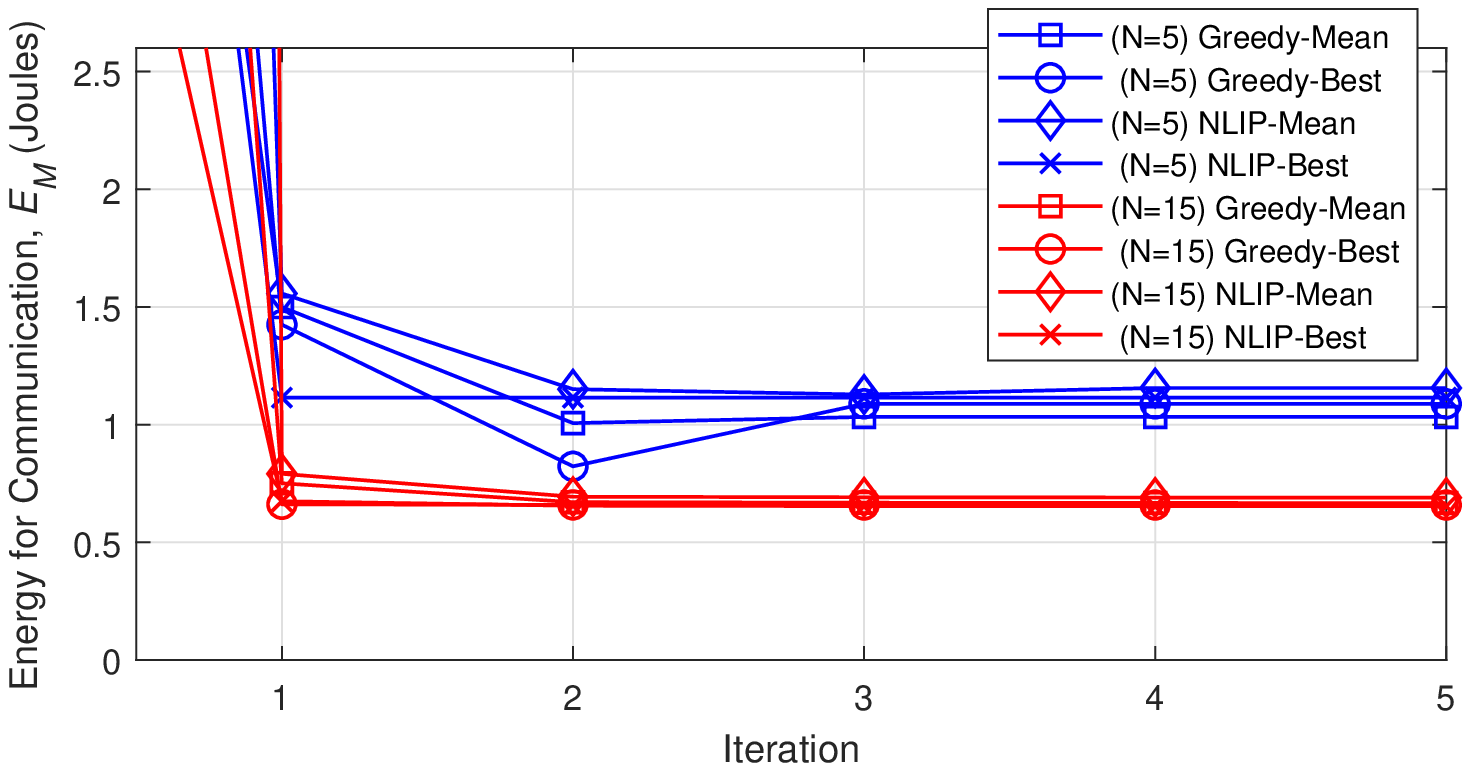}
  \label{fig:iter:EM}
\end{subfigure}
\begin{subfigure}{.48\textwidth}
  \centering
  \includegraphics[width=\linewidth]{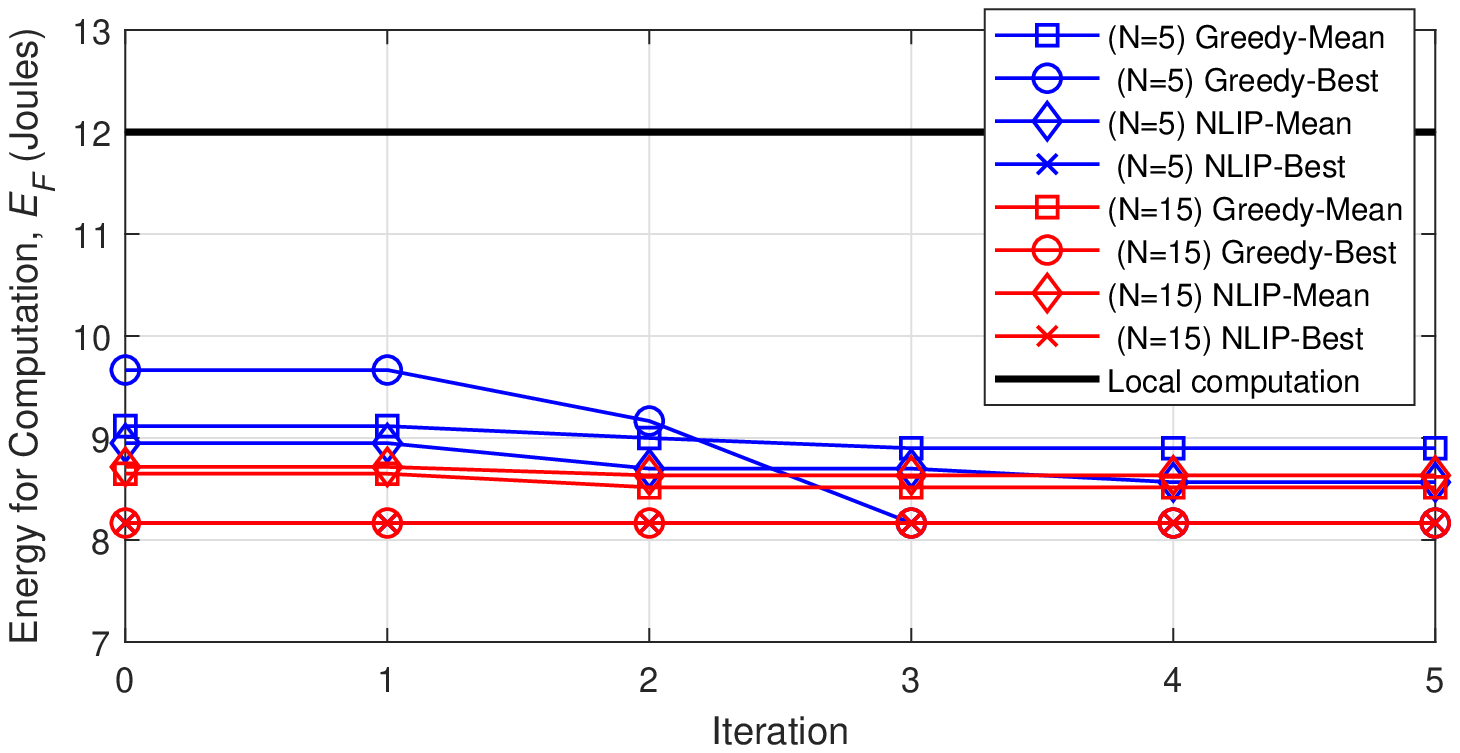}
  \label{fig:iter:EF}
\end{subfigure}
\caption{(a) Total energy $E_P$ (b) Energy for communication $E_M$ (c) Energy for computation $E_F$ along the iteration for two proposed schemes, Greedy--WMMSE and NLIP--WMMSE, under different number of antennas. The total energy $E_P$ is significantly decreased after the first iteration because of 
high reduction of communication energy $E_M$ with MIMO signal design optimization. 
Utilizing more antennas decreases the communication energy $E_M$ further.}
\label{fig:iter}
\end{figure}

Figure \ref{fig:iter} illustrates the convergence behavior of the proposed algorithm along the iteration number under different number of antennas with $K=6$ nodes $S=3$ subchannels, where from the top to bottom, the total energy $E_P$, communication energy $E_M$, and computation energy $E_F$.
The total energy decreases as the iteration proceeds and converges after 2-3 iteration.
After first iteration  \emph{the total energy $E_P$ is greatly decreased because of the sizable reduction in communication}.
This implies that 
both the MIMO signal design and network resource optimization are critical for minimizing the total energy.
Further, the total energy becomes less than that of local computation without offloading, which is denoted by bold line in Figure \ref{fig:iter}.
Also, as the number of antennas $N$ is increasing, from $N=5$ to $N=15$, the interferences from transmit nodes are suppressed further due to the proposed MIMO signal design algorithm for transmit power, beamformer and receive combiner, which leads to less communication energy $E_M$ and total energy $E_P$.


\subsection{Varying number of subchannels}
\label{ssec:sub}

\begin{figure}
\begin{subfigure}{.48\textwidth}
  \centering
  \includegraphics[width=\linewidth]{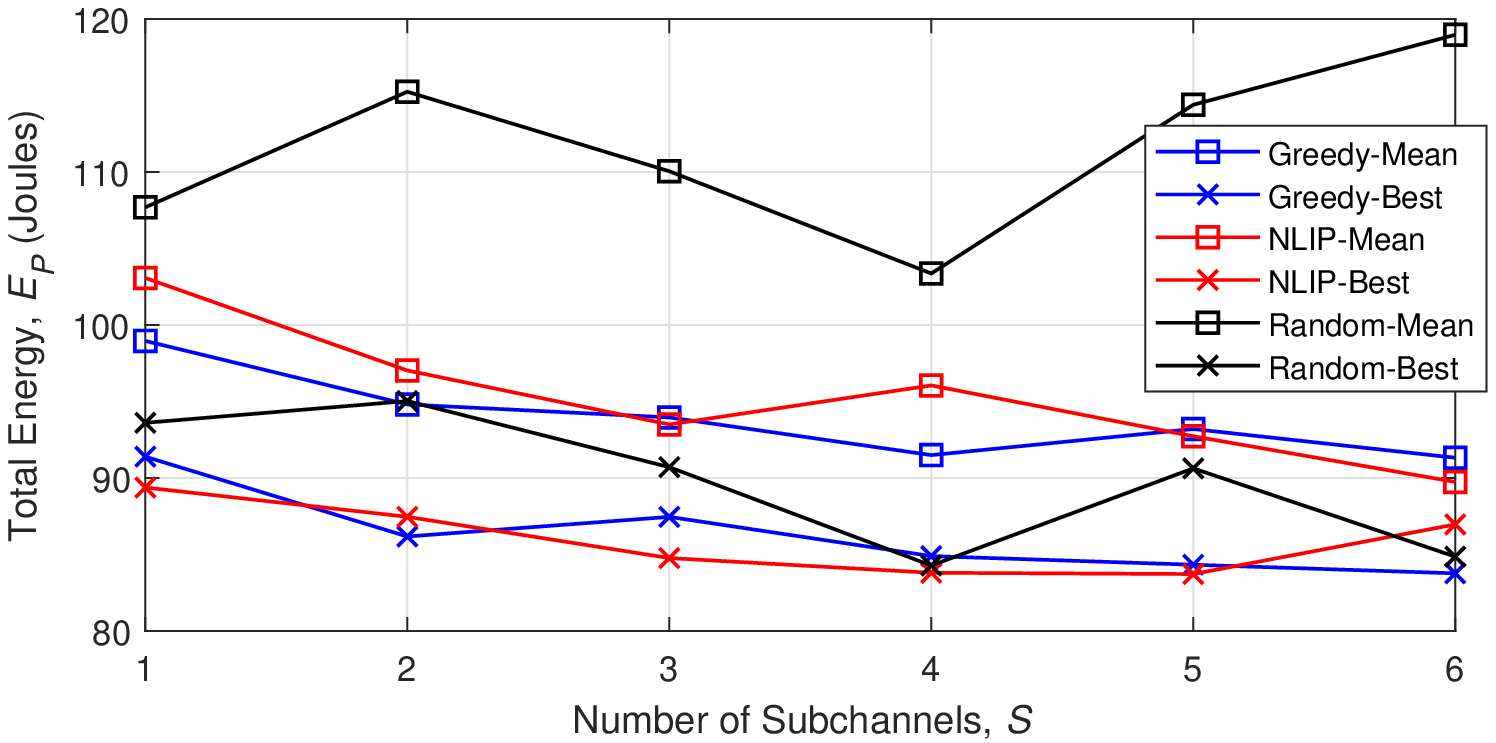}
  \label{fig:sub:EP}
\end{subfigure}
\begin{subfigure}{.48\textwidth}
  \centering
  \includegraphics[width=\linewidth]{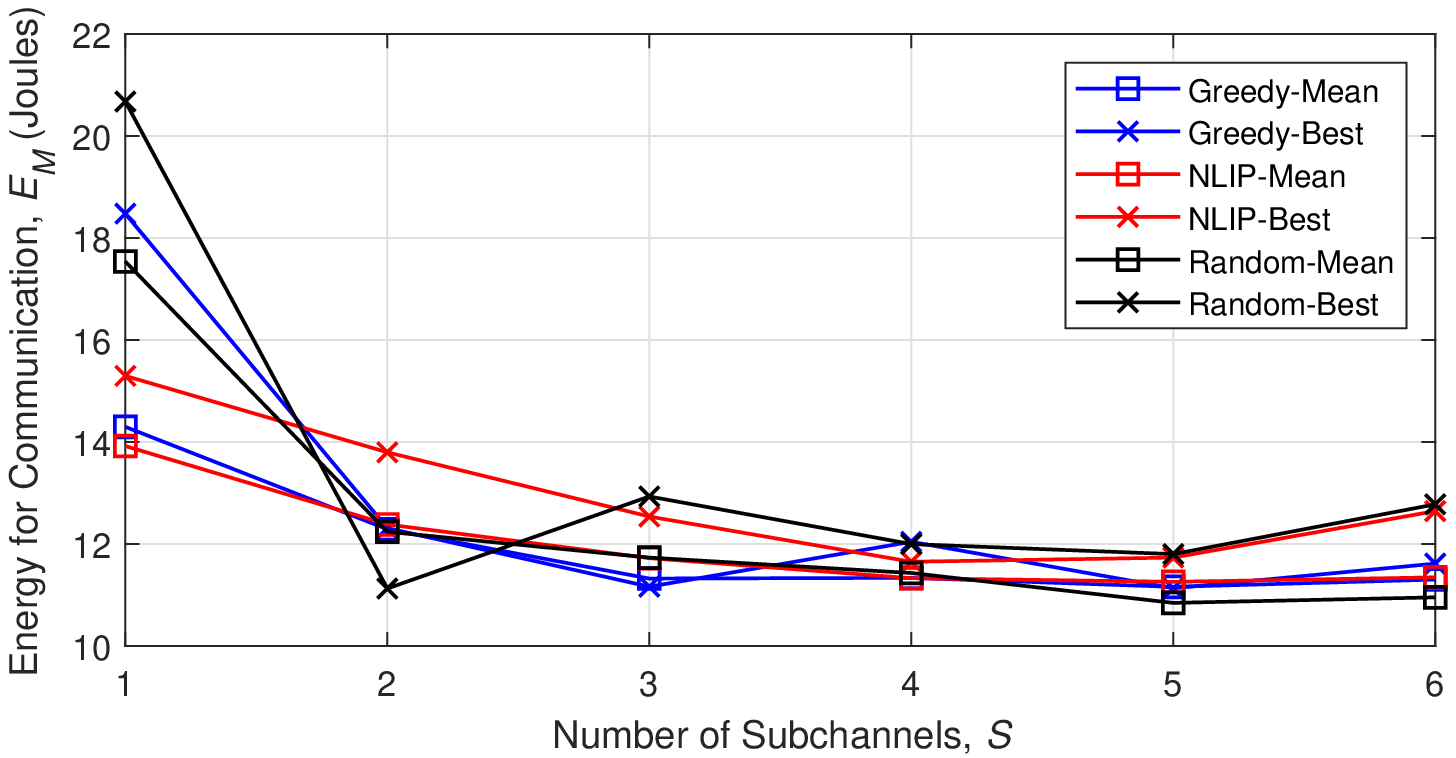}
  \label{fig:sub:EM}
\end{subfigure}
\begin{subfigure}{.48\textwidth}
  \centering
  \includegraphics[width=\linewidth]{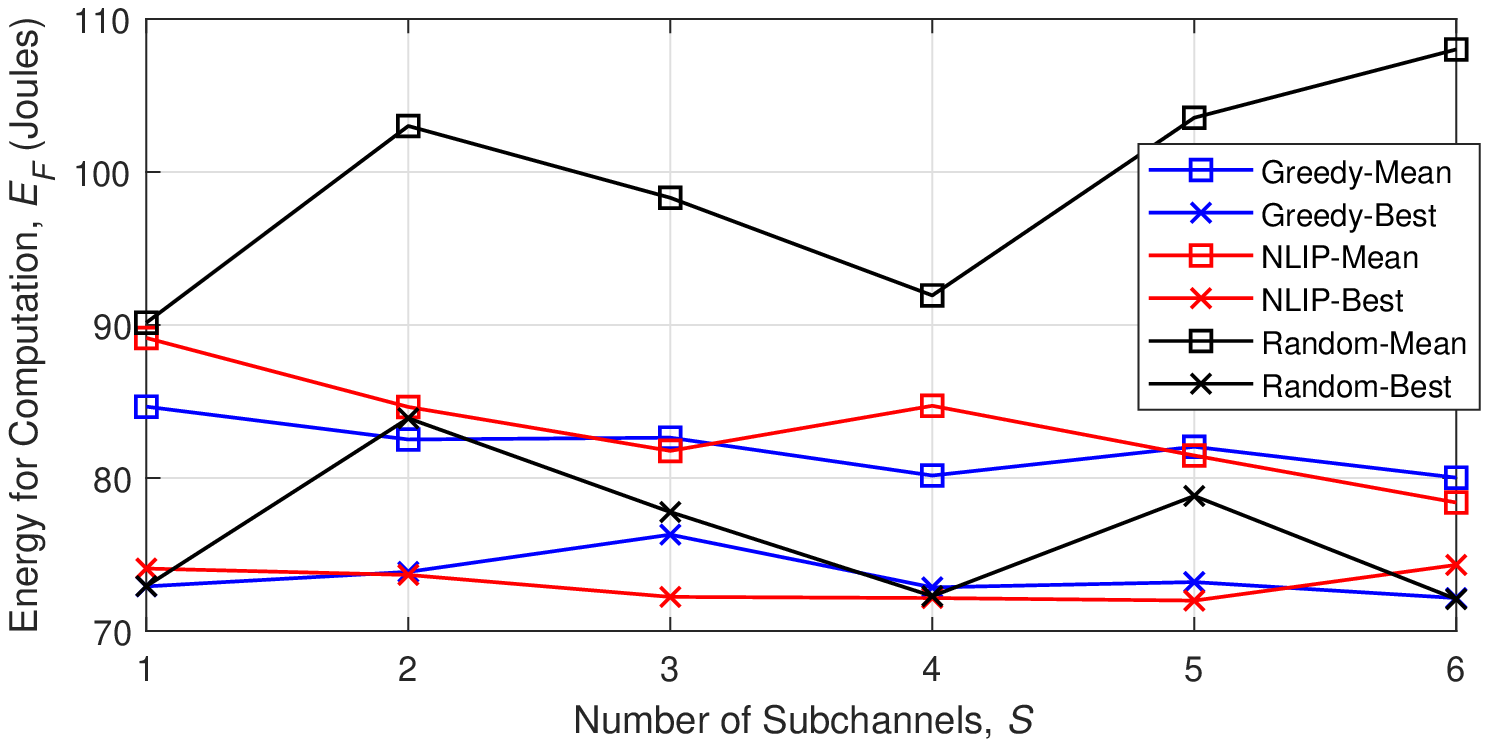}
  \label{fig:sub:EF}
\end{subfigure}
\caption{ (a) Total energy $E_P$ (b) Energy for communication $E_M$ (c) Energy for computation $E_F$ across the number of subchannels $S$ for three different schemes. The total energy $E_P$ decreases as the subchannels increase in number due to the reduction of interferences among nodes.}
\label{fig:sub}
\end{figure}

We consider $K=10$ nodes and $N=7$ antennas for both transmit and receive.
To evaluate joint optimization for MIMO signal design, resource allocation and link selection, we introduce a random matching method, Random--WMMSE which is composed of the random network resource allocation--link selection and subchannel allocation, and MIMO signal design optimization--transmit beamformer with power, and receive combiner with WMMSE approach, i.e., Random--WMMSE is a partially optimized solution.
To be more specific, for Random--WMMSE, 
1) $L_{max} = \lfloor K/2 \rfloor$ link pairs are randomly selected among $K$ nodes and subchannels are allocated to the selected link pairs randomly, 
2) the direction of the links (transmit and receive nodes) is determined by comparing the computation speed $C_k$, 
and 3) the WMMSE approach in Algorithm \ref{al:fz} is exploited for MIMO signal design.
Random--WMMSE method is denoted by Random in Figs. \ref{fig:sub}-\ref{fig:node}.

Figure \ref{fig:sub} compares the energy of the two proposed methods, NLIP--WMMSE and Greedy--WMMSE, with Random--WMMSE across different numbers of subchannels, $S$. 
The proposed methods, NLIP--WMMSE and Greedy--WMMSE, give better performance as compared with Random-WMMSE due to the efficient link allocation to limited number of subchannels. 
{\it The total energy $E_P$ decreases across the number of subchannels}. 
This is a direct consequence from the reduction of $E_M$ and $E_F$.
First, with respect to communication energy $E_M$, more subchannels enable the receive nodes to have less interferences, which increases the data rate and hence decreases communication energy $E_M$.
As the data rates increase, more links are established, which make the computation energy $E_F$ further decreased.
In other words, increasing the ratio $S/K$ enables more reduction of total energy consumption. 
The obtained total energy can fluctuate due to different initializations when the number of subchannels is greater than 3 shown in Figure \ref{fig:sub}. This will be mitigated with much more number of initializations.


\subsection{Varying number of nodes}
\label{ssec:node}

\begin{figure}
\begin{subfigure}{.48\textwidth}
  \centering
  \includegraphics[trim={0 120 0 150},clip,width=\linewidth]{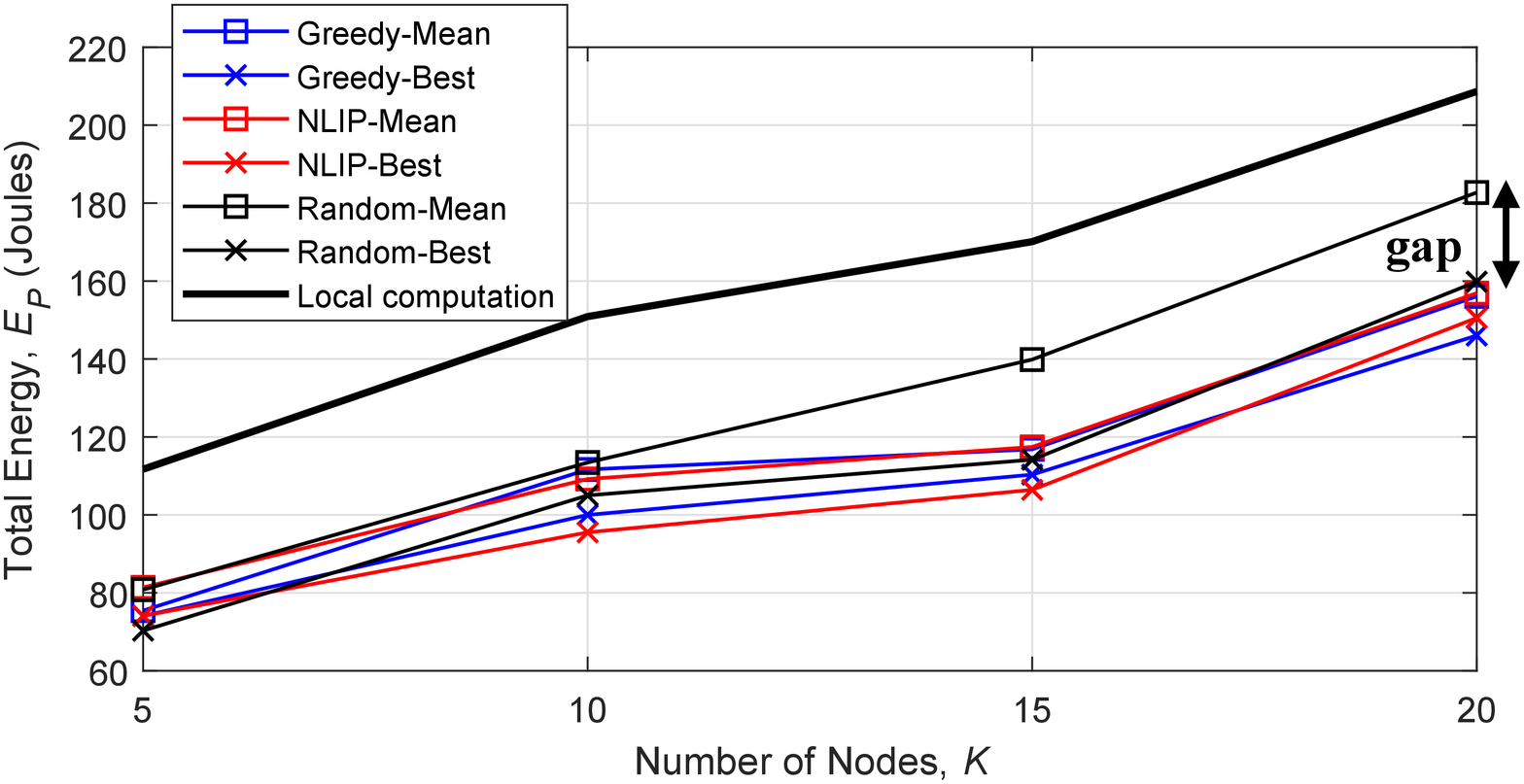}
  \label{fig:node:EP}
\end{subfigure}
\begin{subfigure}{.48\textwidth}
  \centering
  \includegraphics[width=\linewidth]{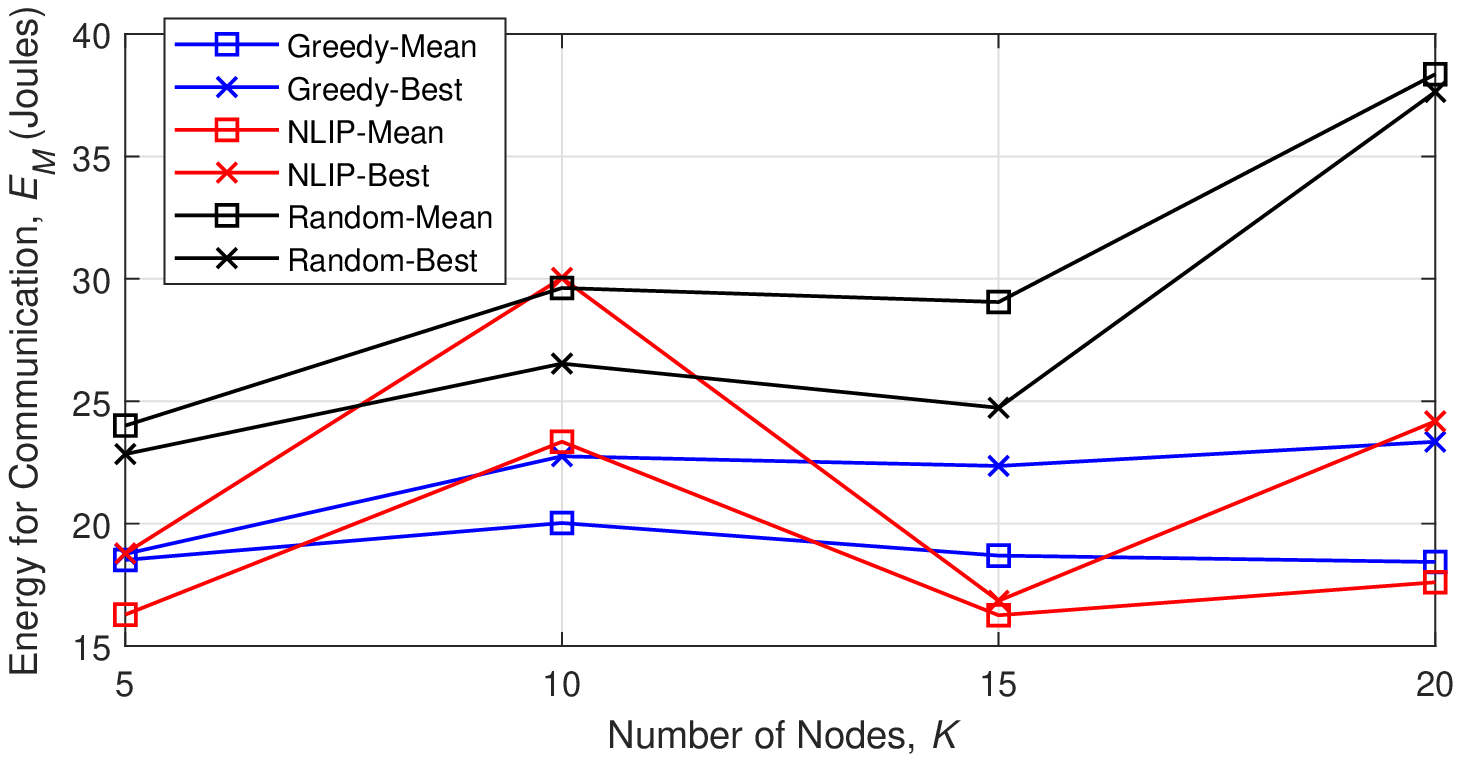}
  \label{fig:node:ET}
\end{subfigure}
\begin{subfigure}{.48\textwidth}
  \centering
  \includegraphics[width=\linewidth]{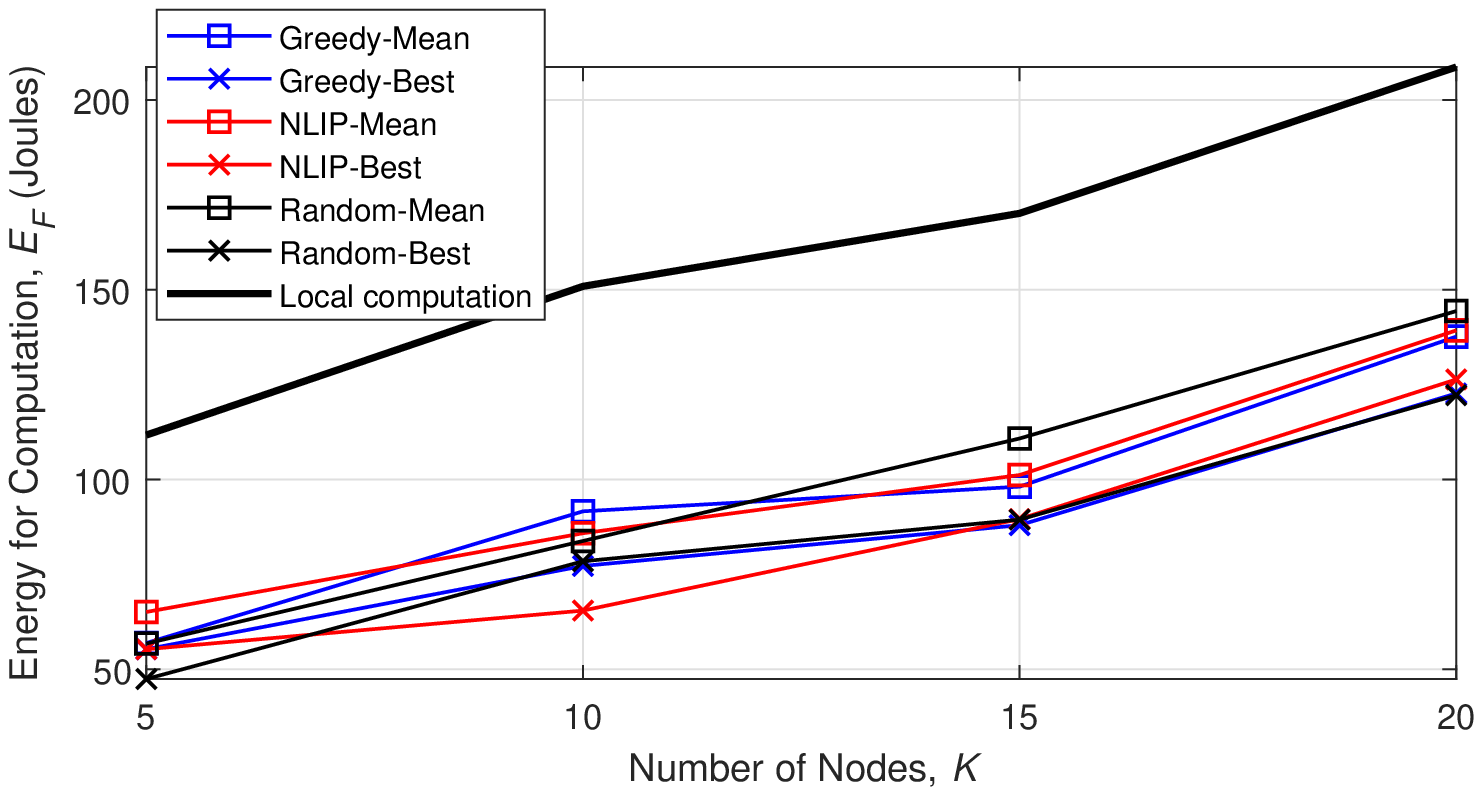}
  \label{fig:node:EF}
\end{subfigure}
\caption{ (a) Total energy $E_P$ (b) Energy for communication $E_M$ (c) Energy for computation $E_F$ comparison of three different schemes along the number of nodes, $K$. 
As the number of nodes increases, the Greedy-WMMSE and NLIP-WMMSE both considering joint optimization shows better performance in minimizing the energy, compared to the partially optimized solution, the Random--WMMSE.
}
\label{fig:node}
\end{figure}

Figure \ref{fig:node} plots the energy $E_P$, $E_M$, and $E_F$ versus number of nodes, $K$. 
We consider $N=10$, $S=2$, and $P=10$.
As the nodes increase in number, it is obvious that the total energy $E_P$ and computation energy $E_F$ are increasing.
However, under constant power constraints $P=10$, the communication energy $E_M$ may not be increasing across the number of nodes because the nodes shares the total power $P$ among nodes.
Interestingly, the Greedy--WMMSE and the NLIP--WMMSE gives almost same result even when $K=20$.
However, {\it the total energy gap between the proposed methods and Random--WMMSE becomes remarkable as the number of nodes is increasing.}
This stresses that efficient, joint MIMO signal design and network resource allocation is critical to maximize the energy efficiecy in the environment surrounded by many devices with different computation speed $C_k$, power $F_k$ and length of data $I_k$.




\section{Conclusion and Future Work}
\label{sec:conc}

In this paper, we proposed a novel optimization methodology for joint MIMO signal design and resource allocation to maximize the energy efficiency of data processing in wireless device-to-device (D2D) edge networks. Given that the problem is a non-convex mixed integer program, we decomposed it into two subproblems for solvability at scale: (1) network resource allocation, i.e., link selection with subchannel allocation, and (2) MIMO signal design, i.e., transmit beamformer, transmit power and receive combiner, and then optimized each problem alternately.
Our evaluation showed that substantial improvements in resource utilization can be achieved through this joint optimization when there are more available link pairs and subchannels for data transfer and interference mitigation. Additionally, we showed that the solution scales well with the size (number of nodes) comprising the network.

%

For future work, time consumption for processing the data can also be included together with energy consumption.
By extension, each receive node can be relaxed to have multiple receive streams.
We are required to design proper receive combiner  and to solve the computing resource allocation problem.
Further work is needed to understand the algorithm behavior, convergence and sub-optimality.


\section*{Acknowledgment}
We thank the reviewers for their valuable comments. D.J. Love was supported in part by the National Science Foundation (NSF) under grants CNS1642982 and CCF1816013.




\end{document}